\documentclass[letterpaper]{article} 
\usepackage{aaai2026}  
\usepackage{times}  
\usepackage{helvet}  
\usepackage{courier}  
\usepackage[hyphens]{url}  
\usepackage{graphicx} 
\usepackage{natbib}  
\usepackage{caption} 
\usepackage{algorithm}
\usepackage{algorithmic}

\usepackage{booktabs}
\usepackage{amsmath}
\usepackage{cleveref}

\usepackage{newfloat}
\usepackage{listings}
\DeclareCaptionStyle{ruled}{labelfont=normalfont,labelsep=colon,strut=off} 
\floatstyle{ruled}
\newfloat{listing}{tb}{lst}{}
\floatname{listing}{Listing}
\nocopyright

\title{L2-Bench: An Evaluation Benchmark for Measuring LLM Capabilities in Second Language Education}
\author{
    James Edgell\equalcontrib\textsuperscript{\rm 1},
    Wm. Matthew Kennedy\equalcontrib\textsuperscript{\rm 2, \rm 3},
    Ben Knight\textsuperscript{\rm 1},\\
    Danielle Carvalho\textsuperscript{\rm 1},
    Martin Ku\textsuperscript{\rm 1},
    Isaac Pattis\textsuperscript{\rm 1}
}
\affiliations{
    \textsuperscript{\rm 1}Oxford University Press\\
    \textsuperscript{\rm 2}Oxford Internet Institute,  University of Oxford\\
    \textsuperscript{\rm 3}King's College London\\
    elt-bench@oup.com
}

\usepackage{bibentry}

\begin{document}

\maketitle

\begin{abstract}
Despite rapid AI adoption in education, rigorous evaluation of AI-powered educational (AIED) systems remains critically underdeveloped, particularly in second language (L2) education, one of the most common yet least evaluated AI applications. We introduce L2-Bench, an open-source benchmark of 1,000+ task-response pairs to aid the pedagogy-led evaluation of LLM capabilities relating to language learning and assessment. Crucially, L2-Bench measures model performativity on the application of learning experience design principles rather than mere knowledge of those principles or broad learning outcomes. Our contributions include: (1) a validated taxonomy of 12 competencies and 31 subcompetencies validated by 200+ expert practitioners (task authenticity = 4.42/5.00, criteria adequacy = 4.18/5.00); (2) a rubric-based evaluation methodology that we believe can, if adapted, generalize to similar (open-ended, qualitative) disciplines; (3) an evaluation dataset that produces reliable signal about model strengths, weaknesses, and contextual robustness across diverse L2 education scenarios. We find that, among large models, Claude Opus 4.7 performs best overall (85.5\%), though is marginally outperformed on several constituent tasks. We also find that performance drops notably on harder tasks $\in$ (69.9\%–73.4\%). L2-Bench provides education stakeholders better methods to make more informed decisions about real-world AIED adoption, use, and governance, while advancing the maturing science of AI evaluations for education.
 
\end{abstract}


\section{Introduction}
\label{sec:intro}

Despite rapid adoption of AI systems in educational spaces \cite{digitaleducationcouncil2024globalai}, very few evaluations for AI in educational (AIED) exist. Those that purport to cover this space often suffer from poor construct validity, are underpowered, or focus more on broad effects rather than instance-level performativity. The result is little short of a “wild, wild west”: a situation in which deployment is far outpacing evidence even of basic validation of AIED systems \cite{10.1145/3772318.3791968}.  

This AIED evaluations gap reflects the ongoing crisis in AI evaluations in general. However, it is more acute: although prior validation and efficacy studies of conventional education technologies suggest that these predecessor systems have not meaningfully improved learning outcomes \cite{cuban2001oversold, zawacki2019systematic, selwyn2012education, Selwyn2019Robots}, many AIED systems replicate or extend the same conventional system design patterns \cite{Jurenka2024ResponsibleEduAI}. Such a headlong rush on the part of AIED developers and adopters without parallel development of novel evaluation methodologies creates the conditions for myriad interaction-, learning-group-, systemic-, and compounding harms \cite{Bastani2024HarmLearning, HolmesMiao2023UNESCO, KASNECI2023102274, Wachter2024Truth, Holmes2024AIED, 10.5555/3716662.3716723}. In the meantime, a generation of learners must endure laboratory conditions \cite{alcaras2025configurationworkconsequencesllmsinuse} in their pursuit of educational attainment.

Drawing inspiration from allied efforts to improve the state of the evaluations ecosystem \cite{reuel2024betterbench, eriksson2025trust, biderman2024lessonstrenchesreproducibleevaluation, weidinger2025evaluation, schwartz2025realitychecknewevaluation, bean2025measuring, paskov2025toward, reuel2025evaluatesaissocialimpacts}, our work hastens progress in one particularly challenging subdomain of AI for education: second language (L2) education. The use of LLMs for language learning tasks represents one of the most common applications of AI models \cite{Tamkin2024Clio, costagomes2025copilot, doi:10.1177/00336882231162868}, yet it is among the least evaluated. To work towards closing this gap, we propose an evaluation benchmark for \textbf{second language learning experience design} for L2 US/UK English. Our benchmark makes three distinct contributions: 

\textbf{1)} It provides a validated taxonomy of the 12 competencies and 31 subcompetencies that compose second language learning design, one which operationalizes discipline-specific practices and frameworks into a sociotechnical boundary object that promotes the uptake of domain knowledge into current AI systems design. A large sample of n = 221 expert practitioner raters scored task authenticity and criteria adequacy highly (task authenticity = 4.42/5.00, criteria adequacy = 4.18/5.00).

\textbf{2)} It offers an evaluation methodology for assessing model performance across each of these competences. We advance, as a research agenda item rather than a validated claim, the conjecture that this rubric-based methodology could---with substantial domain adaptation---be extended to other educational domains that also feature well-developed assessment and pedagogical frameworks within SHAPE disciplines (i.e. the social sciences and humanities); establishing this generalization empirically is left to future work and discussed in Appendix~D.

\textbf{3)} It produces a benchmark dataset of 1000+ task-response item pairs structured by our taxonomy, ultimately allowing stakeholders as diverse as instructors, administrators, policymakers, learning scientists, AI evaluations scientists, AI application developers, and AI model developers to better understand the current and near-future state of AI model capabilities in this “high risk” social sector.

We evaluate nine popular models, finding that, overall, among frontier models, Claude Opus 4.7 performs best on benchmark tasks (85.5\%), but is marginally less performant in other areas (managing activities, presenting linguistic points, or student assessment). On hard tasks, large model performance falls to $\in$ [69.9\%, 73.4\%], suggesting that there is room for improvement—and good reason to develop more complex evaluation datasets. 

We propose this benchmark as the first offering of our wider program to develop context-specific AIED evaluation methodologies. As such, L2-Bench prioritizes essential desiderata that second language learning design benchmarking should seek first to capture. We also provide insights into new directions for AIED evaluations methods in general.


\section{The AIED evaluation problem}
\label{sec:eval-problem}

Education is a domain constituted by its own distinctive values, norms, and discourses \cite{Jacobson01012006, Weidinger2023Sociotechnical, Kennedy2026Vernacularized}, ones that do not map cleanly onto the performance metrics native to current approaches to AI evaluation \cite{biesta2010good, luckin2016intelligence}. Education stakeholders themselves disagree about how to measure the realization of such values \cite{pellegrino2012education, 10.1145/2883851.2883893}. This makes for an evaluations challenge. Recent work broadly clusters into three categories: evaluations of precursor model capabilities, evaluations of outcomes and systemic impacts, and evaluations of model performance on highly domain-specific tasks or processes.

\textbf{Precursor capability evaluations} assess whether a model possesses the underlying capabilities presumed to be necessary for downstream use in certain domains \cite{brown2025precursors}. For education, these might include instruction-following \cite{zhou2023instructionfollowingevaluationlargelanguage}, question answering \cite{rajpurkar2016squad, yang2018hotpotqa}, logical reasoning \cite{clark2021fanyregents, tafjord2021proofwriter}, reading comprehension \cite{kocisky2018narrativeqa, pang2022quality}, mathematical problem-solving \cite{cobbe2021trainingverifierssolvemath, hendrycks2021measuringmassivemultitasklanguage}, and knowledge of domain content and reasoning/cognitive-process \cite{Krathwohl01112002}. 

\paragraph{General pedagogical capability} is perhaps the most important precursor capability for AIED evaluations. The most sustained evaluation effort to date is Google DeepMind's LearnLM program \cite{Jurenka2024ResponsibleEduAI}, which is focused on pedagogical behavior in tutorial interaction contexts. However, LearnLM’s evaluations remain proprietary, and its focus on tutorial interaction privileges instruction-following over holistic pedagogical adaptivity \cite{bordes2025evalfactsheetsstructuredframework, https://doi.org/10.1111/j.1551-6709.2012.01245.x}. Additionally, Xu et al \shortcite{xu2026edubenchcomprehensivebenchmarkingdataset} propose EduBench, a general-purpose evaluation framework, although it lacks grounding in pedagogical theory. Clark et al \shortcite{clark2025autoevaluation} offer Oak Academy’s useful safety-focused benchmark for lesson generations, but its scope does not extend to pedagogical quality or learner outcomes. Lelièvre et al \shortcite{lelièvre2025benchmarkingpedagogicalknowledgelarge} propose a benchmark for LLM knowledge of pedagogical concepts, but not their application. Shetye's \shortcite{shetye2024khanmigo} analysis of Khanmigo, although well-grounded in Chapelle’s \shortcite{chapelle2001computer} computer-assisted language learning (CALL) framework, relies only on anecdotal experience. Maurya et al \shortcite{maurya2025unifying} implemented their team’s prior taxonomy of LLM tutor pedagogical capabilities to evaluate AI tutorial chatbot performance on identifying the nature and location of student mistakes in tutorial interactions, on provision of guidance, and the feasibility of feedback offered. Shi et al.’s \shortcite{shi2025educationq} EducationQ evaluated several leading LLMs on a variety of (synthetic) instructional tasks, finding that model size does not correlate to pedagogical performativity. EduEVAL-DB \cite{irigoyen2026eduevaldbrolebaseddatasetpedagogical} enables the evaluation of LLM tutor instructional explanations across five criteria, including pedagogical risk, although the dataset is limited to responses to 139 questions from ScienceQA \cite{lu2022learn}, only one-sixth of which are produced by human experts. Kennedy and Campos \shortcite{Kennedy2026Vernacularized} advance a vernacularized taxonomy of harms for AIED, however it has yet to be operationalized into benchmark infrastructure.  

\paragraph{Outcome and systemic evaluations} measure the effects of AI system deployments to specific sectors or actors within those sectors. For education, this might include instructors, learners, and educational systems more broadly. Effects might range from teaching efficiencies, learning gains, engagement, and equity of access among others \cite{holmes2022ethics, educsci13070692, unesco2025ai}. Where controlled studies exist, effect sizes are modest and generalizability is sometimes limited by inauthentically scoped task definitions as well as the rapid obsolescence of the systems under evaluation \cite{doi:10.3102/0034654315581420, zawacki2019systematic}. A growing body of scholarship suggests AI use may yield marginal learning benefits but also influences (sometimes negatively) engagement \cite{friedman2026shortlongllmresponse, morris2026feedbackdifferentsourceai}, learner autonomy \cite{furuhashi2026feedbackworkswhomdifferential, borchers2026decidesaimediatedlearningagency}, and deeper cognitive processing \cite{pardos2023learninggaindifferenceschatgpt, 10.1145/3698205.3733960, Bastani2024HarmLearning, HADIMOGAVI2024100027, 10.1016/j.chb.2024.108386, hu2025teaching, gao2026doubao}. Similar blinded studies exploring LLM-generated teaching tasks \cite{PhysRevPhysEducRes.19.020128} or K-12 lesson plans suggest LLMs can produce teaching materials of comparable quality to human-produced materials, but this does not necessarily entail workflow efficiencies. The Google DeepMind LearnLM team’s recent partnership RCT with Eedi (N = 165) reported a 5.5\% improvement in independent problem-solving over human tutoring alone \cite{learnlmteam2025aitutoringsafelyeffectively}, but a single industry-conducted trial of this scale is insufficient to establish the evidentiary foundation that broad deployment requires \cite{cook1979quasi, Slavin03072017, Selwyn2022Future, unesco2025ai}.

\paragraph{Domain-specific task evaluations} assess model performance on domain-specific tasks that are meaningfully educational in character \cite{Jurenka2024ResponsibleEduAI, 2024.EDM-short-papers.49, edgell2026accuracyrobustevaluationmethodology}, such as giving formative feedback on essays \cite{10.1145/3772318.3790539}, cybersecurity lesson or curriculum planning \cite{nijdam2026curricullmdesigningpersonalizedworkforcealigned}, generating science practice problems at a specified difficulty level \cite{brant2026estimatingexamitemdifficulty, hatchett2026learningcontextmattersmeasuring}, or identifying potential misconceptions in student responses \cite{zengaffinen2026llmsmodelincorrectstudent}. 

The overwhelming majority of domain-specific AIED evaluations have been conducted in highly structured educational domains. These include dialogue-based mathematical tutoring \cite{macina2023mathdial} and error correction \cite{10.1007/978-3-031-36272-9_30}. Other benchmarks evaluate scientific question answering across difficulty levels \cite{welbl2017crowdsourcing, clark2018thinksolvedquestionanswering, lu2022learn, sun2024scievalmultilevellargelanguage, wang2024scibench}. In computer science and programming, HumanEval \cite{chen2021evaluatinglargelanguagemodels} and its successors DS-1000 \cite{lai2022ds1000naturalreliablebenchmark}, and LiveCodeBench \cite{jain2024livecodebenchholisticcontaminationfree} provide automated evaluation via test-case execution, while CS1QA \cite{lee2022cs1qa} and CSEDM datasets (among others) evaluate the pedagogical appropriacy of AI feedback on novice programming work. 

Evaluations methods for more determinate domains have proven brittle when imported into open-ended domains. In writing assessment, for instance, ASAP \cite{shermis2013handbook} and PERSUADE \cite{CROSSLEY2022100667} provide automated essay scoring datasets, but these only measure surface features rather than argumentation quality, rhetorical effectiveness, disciplinary thinking \cite{article, chapelle2008toefl}. Feedback quality benchmarks such as FeedbackQA \cite{Li_2022} and those introduced in the context of automated writing evaluation \cite{beigman-klebanov2020automated} rely on costly human annotation, preventing scaling and amplifying rater disagreement challenges (which, of course, is inherently higher in less structured domains). Progress here is being made, however. Du et al.'s \shortcite{du2026benchmarkingeducationalllmsanalytics} recent benchmark for bias in essay-writing feedback is a notable empirical and methodological contribution to the emerging subfield, even if focused narrowly on feedback quality.

\paragraph{AI for language education} entails yet another challenge: natural language is simultaneously the target and the medium of learning \cite{cook1992discourse, larsenfreeman2003teaching}. Language education is distinctive in requiring largely implicit, proceduralised knowledge \cite{dekeyser2007practice, inbook}, but, in execution, is heavily shaped by affective factors including motivation, identity, and anxiety \cite{https://doi.org/10.1111/j.1540-4781.1998.tb05543.x, papi2023second, dornyei2009psychology}. Language is more than just “language data” \cite{smart2024sociallyresponsibledatalarge}. Its effective use is codetermined by each learner's social and cultural experience in ways that resist standardization \cite{10.1093/applin/amm027, norton2013identity, block2007second}. As language is never "solved” \cite{inbook, ortega2009understanding}, language education has proved very difficult to measure. 

Existing AIED evaluation efforts reflect these difficulties. Benchmarks and studies targeting grammatical error correction \cite{bryant2019bea, ng2014conll}, reading comprehension \cite{xie2018large}, essay-writing \cite{shermis2013handbook, 10.1108/979-8-88730-606-320251003}, and vocabulary \cite{pilan2016coursebook} target isolated sub-skills rather than integrated communicative competence \cite{bachman1990fundamental, canale1980theoretical}. More educationally grounded resources like the Cambridge Learner Corpus \cite{Nicholls1999TheCL}, EFCAMDAT \cite{Geertzen2014AutomaticLA}, and ICNALE \cite{ishikawa2013icnale} provide authentic learner language but were not designed to evaluate LLMs, and none adequately operationalises uptake \cite{Lyster_Ranta_1997}, interactional scaffolding \cite{10.1093/applin/amm027}, or appropriacy \cite{BardoviHarlig2001PragmaticsIL}.

Only a handful of evaluation benchmarks in this subdomain exist. Existing AI-specific language learning evaluations, including studies of chatbot tutoring \cite{doi:10.1177/00336882231162868, godwinjones2022partnering}, automated writing evaluation \cite{Ranalli03092018, Stevenson_Phakiti_2019}, and speaking assessment \cite{https://doi.org/10.1002/ets2.12080}, offer partial coverage but fail to engage established pedagogical frameworks or demonstrate construct validity. Those that do primarily target LLM performance on L2 (English) automated essay scoring relative to a prior generation of conventional auto-scoring systems \cite{MIZUMOTO2023100050, yancey2023rating} and frequently focus only on the student-instructor-material triad, neglecting social and affective dimensions. Meyer et al.'s \shortcite{MEYER2024100199} large RCT (N = 459) found significant (d = 0.19) effects of LLM-generated L2 English essay writing feedback on revision performance, but only in comparison to no feedback whatsoever. In most other dimensions, the field is silent.


\section{Presenting L2-Bench}
\label{sec:l2bench}

We present L2-Bench, the first evaluation benchmark to assess LLM second language learning design capabilities. L2-Bench includes a novel taxonomy comprising 12 core competencies and 31 sub-competencies; an evaluation dataset of 1000+ expert-reviewed and fully validated task-response pairs; and a scoring pipeline that provides for automated measurement using an optimized LLM-as-a-judge. At this current time, L2-Bench assesses learning experience design in UK/US L2 English. We focus on \textit{UK/US English first} because it is the most-studied second language in 79\% of all countries—the evaluation need is greatest here \cite{blanco2025duolingo}.

\subsection{Representing second language learning design as a construct}
\label{subsec:construct}

L2-Bench specifically targets a construct expressed as “second language learning experience design.” We ask: to what extent are LLMs capable of supporting second language learning design tasks? We consider the question of efficacy—whether LLM-assisted second language education produces better learning outcomes—out of scope.  

Evaluation for learning design capabilities, rather than instruction, better reflects the breadth of real-world use of AIED for language education \cite{Tamkin2024Clio, costagomes2025copilot}. Instructors use LLMs in L2 educational contexts to support a wide variety of tasks. Teaching is one of those tasks, but education is much more than "just teaching,” and teaching is more than mechanistic information processing \cite{HolmesBialikFadel2019, Kennedy2026Vernacularized, edgell2026accuracyrobustevaluationmethodology}. It is also socialization, care, security, and “subjectification” as Biesta \shortcite{Biesta2015} maintains. Each of these dynamics influences how appropriate instruction might proceed, and, consequently, how learning occurs. This also better accounts for the variable temporalities of education. Learning does not happen immediately, but rather proceeds unevenly, unpredictably, and at different rates depending on experience with learning itself \cite{kalyuga2007expertise}. Instructional interventions may not appear to have any immediate impact, yet, after a period of time, can show great effect \cite{bjork2011desirable}. We elaborate in Appendix~D.

\subsubsection{Taxonomy}  
We define the L2 learning design construct as a novel taxonomy comprising 31 subcompetencies grouped into 12 main competencies (full taxonomy in Appendix~E). Competencies and subcompetencies were identified initially via a structured review of the five most influential and high-adopted second language education frameworks (the CEFR \cite{coe2018cefr}, the Eaquals Framework \cite{Eaquals2016Framework}, the Cambridge English Teaching Framework (CETF) \cite{cambridgeenglish_teaching_framework}, the Competency Framework for Language Learning Materials Writing \cite{millin2023competency}, and the British Council CPD Framework \cite{BritishCouncil2025CPD}). Although these frameworks originated in European contexts, they are now widely used and adapted globally, and the CEFR in particular functions as an international reference point across many languages and educational systems. The taxonomy reflects principles of practice that have been stabilized through global uptake rather than norms confined to a single regional context (see also Appendix~D). Competencies and subcompetencies were further refined through expert iteration and findings from a pilot validation study (reported in \cite{edgell2026accuracyrobustevaluationmethodology}).

\subsubsection{Dataset}
\label{subsec:dataset}

To operationalize this taxonomy, we produced an evaluation dataset of 1,000 single-turn task-response pairs. Following the UK AISI's \shortcite{aisi2024insights} and Bahari et al's \shortcite{bahari2025call} framework for AI Computer Assisted Language Learning (AICALL), tasks are designed to be (i) relevant to authentic professional practice, (ii) representative of multiple stakeholder perspectives (teachers, learners, materials developers, assessment specialists), (iii) clear, with sufficient contextual guidance for response generation, and (iv) original---situated in novel combinations of context factors unlikely to have been memorized during pretraining.

Each task is parameterized by context factors drawn from a structured ontology of 33 variables across five dimensions: learner characteristics (age group, CEFR proficiency A1--C2, L1 script, learning needs), learning purpose (curriculum focus, skill focus), learning context (setting type, class configuration, delivery mode, scheduling), teacher factors (proficiency, experience, L1 relationship), and available resources (materials, technology, economic context). This combinatorial structure ensures that tasks span the diversity of real-world L2 educational settings, and also allows for systematic application of rubric criteria, outlined below (see also Appendix~E for context factor details).

Tasks were produced via a hybrid human-AI authoring pipeline modeled on professional publishing workflows: design, draft, expert review, revision, and approval. Items were generated in batches of $\leq$144 (12 per competency), with expert feedback from each batch informing subsequent generation. This iterative process enabled criteria drift detection and progressive quality improvement (see Appendix~F for full production methodology).

\subsubsection{Rubrics}

L2-Bench is a rubric-based benchmark. Rubrics indicate what an ideal response to any given task should entail (or should avoid). Instead of employing Likert scales, we opt for simple pass/fail classifications in order to better calibrate human reviewers and LLM judges and to promote clearer annotation signal \cite{yan2024llmevaluators}. We assign point values $\in$ $[-10, +10]$ to criteria based on importance, and award negative point values for undesirable responses. A task's final score is the weighted sum of passed criteria divided by the sum of positive weights only (negative criteria act as penalties; see Appendix~I).

We implement a three-layer criteria system (see the Task Criteria Design subsection of Appendix~F) to account for the hierarchical relationship between sub-competencies, individual tasks, and domain-wide considerations. Reference answers are provided for each task to guide both human and LLM scoring.

Consensus criteria encode the shared expectations that expert practitioners hold for responses within a given sub-competency, ensuring that domain knowledge is systematically represented across all tasks tagged to that sub-competency. Task criteria capture context-specific requirements that emerge from the particular instructional scenario---these are generated after task authoring and are designed to be independent of consensus criteria to avoid score inflation. Universal criteria enforce domain-wide constraints (age appropriateness, CEFR level, cultural sensitivity, data privacy) whose weightings vary according to task context variables (see Appendix~E). The full taxonomy of 12 competencies, 31 sub-competencies, and their associated consensus criteria appears in Appendix~E; worked task examples with all three criteria layers appear in Appendix~F.

\subsection{L2-Bench construct validation}
\label{subsec:validation}

Building on a prior pilot validation exercise \citep{edgell2026accuracyrobustevaluationmethodology}, we designed a large, representative study to validate L2-Bench at scale with domain experts (full study design reported in Appendix~G). The study involved N = 221 practitioners across 45 countries spanning 6 stakeholder groups (content developers, assessment specialists, teachers, generalist education professionals, academics, and learners). We excluded 20 raters who exhibited signals of systematic straight-lining or cheating, resulting in the N = 221 practitioners number quoted above, who yielded 1,447 ratings across 474 items. Overall both task authenticity (M = 4.42, 95\% CI [4.38, 4.46]) and criteria adequacy (M = 4.18, 95\% CI [4.14, 4.22]) significantly exceeded their targets of 4.0 and 3.5 respectively. Both measures met or exceeded targets across all 12 competencies individually, providing strong evidence that the construct is coherent at the sub-competency level.

Although inter-annotator agreement (IAA, via Krippendorff's $\alpha$) did not show strong agreement, this is consistent with studies with sparse coverage ($\sim$3 raters/item on average in our study) and similar evaluation studies in subjective domains \cite{he2026judgingjudgeshumanvalidation}. There is however some evidence of reliability in ratings when considering (a) internal item consistency (IIC, via Cronbach's $\alpha$) for both task authenticity ($\alpha$ = 0.53) and criteria adequacy ($\alpha$ = 0.61) indicated moderate item coherence, and (b) mixed-effects modeling revealed that 27–33\% \ of the rating variance is attributable to stable severity differences at the rater-level (see Appendix~G for details). Recall that while IAA penalizes raw score differences and is sensitive to constant offsets between raters, IIC is mathematically invariant to such offsets, and mixed-effects modeling indicates the presence of systematic per-rater offsets in scale usage, therefore practitioners are applying different pedagogical frameworks in a subjective domain like ours. Ultimately, with individual ratings carrying noise, we therefore conclude that aggregate (competency-level) analysis is more informative than item-level conclusions.

\begin{table*}[h]
\centering
\caption{Model performance ranking on L2-Bench with 95\% confidence intervals ($N = 1{,}000$ tasks per model).}
\label{tab:model_ranking}
\begin{tabular}{l l c c c c}
\toprule
Rank & Model & Score (\%) & SE & CI Lower (\%) & CI Upper (\%) \\
\midrule
1 & Opus 4.7         & 85.5 & 0.0034 & 84.9 & 86.2 \\
2 & GPT 5.4          & 84.1 & 0.0035 & 83.4 & 84.8 \\
3 & Gemini 3.1 Pro       & 83.4 & 0.0037 & 82.7 & 84.2 \\
4 & Gemini 3 Flash     & 80.7 & 0.0038 & 79.9 & 81.4 \\
5 & DeepSeek V3.2    & 80.2 & 0.0042 & 79.4 & 81.1 \\
6 & Kimi K2.5        & 79.1 & 0.0042 & 78.3 & 79.9 \\
7 & Haiku 4.5        & 78.8 & 0.0045 & 77.9 & 79.7 \\
8 & Qwen3 32B        & 65.8 & 0.0063 & 64.5 & 67.0 \\
9 & Magistral Small  & 50.7 & 0.0069 & 49.4 & 52.1 \\
\bottomrule
\end{tabular}
\end{table*}

\begin{figure*}
    \centering
    \includegraphics[width=1\linewidth]{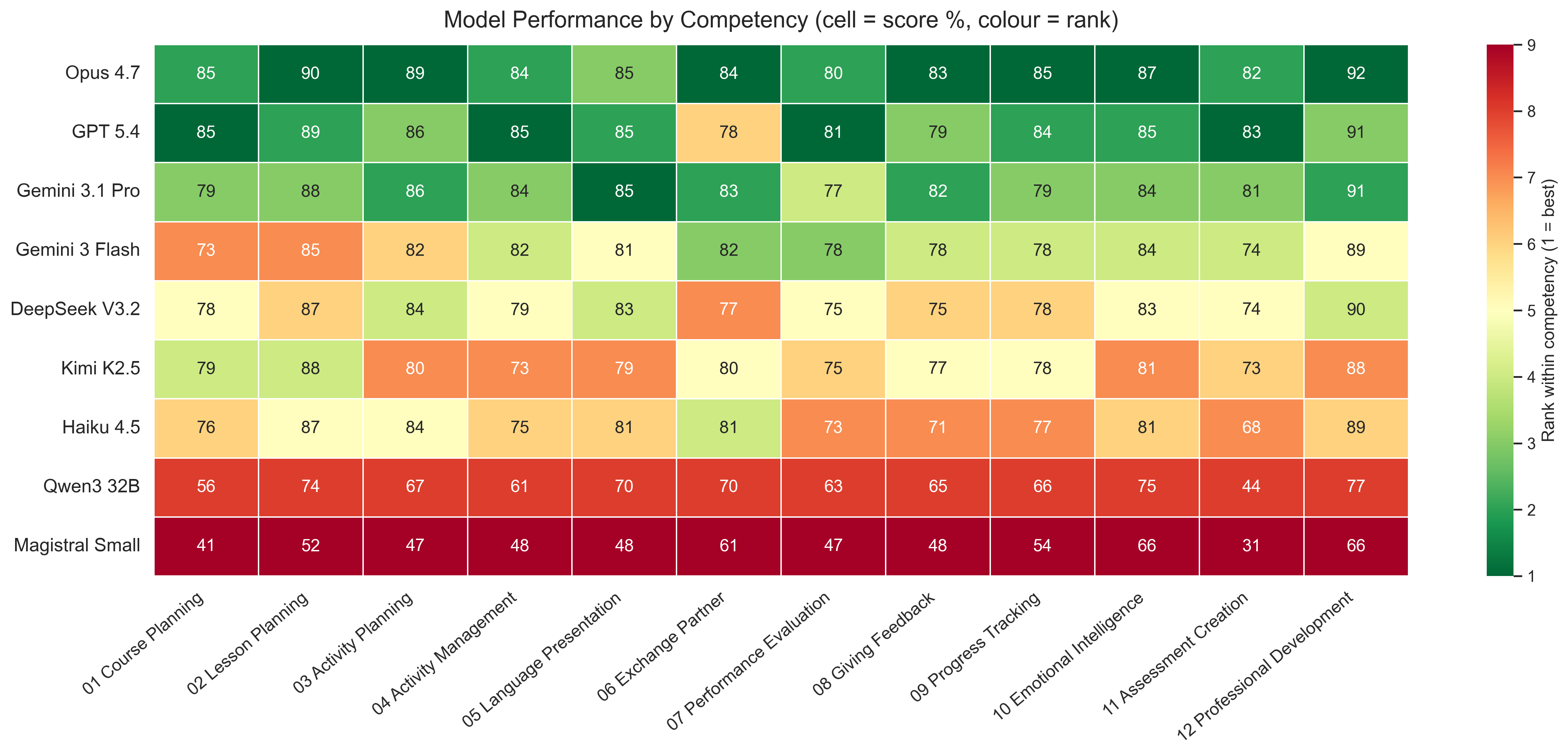}
    \caption{Model performance by competency. Relative strengths and weaknesses are legible independently of a model's overall level. All models perform relatively well on structured output tasks (lesson and activity planning) but less well on open-ended competencies (conversational exchange partner, giving feedback, evaluating performance).}
    \label{fig:model-performance-by-competency}
\end{figure*}
\begin{figure*}[h]
    \centering
    \includegraphics[width=1\linewidth]{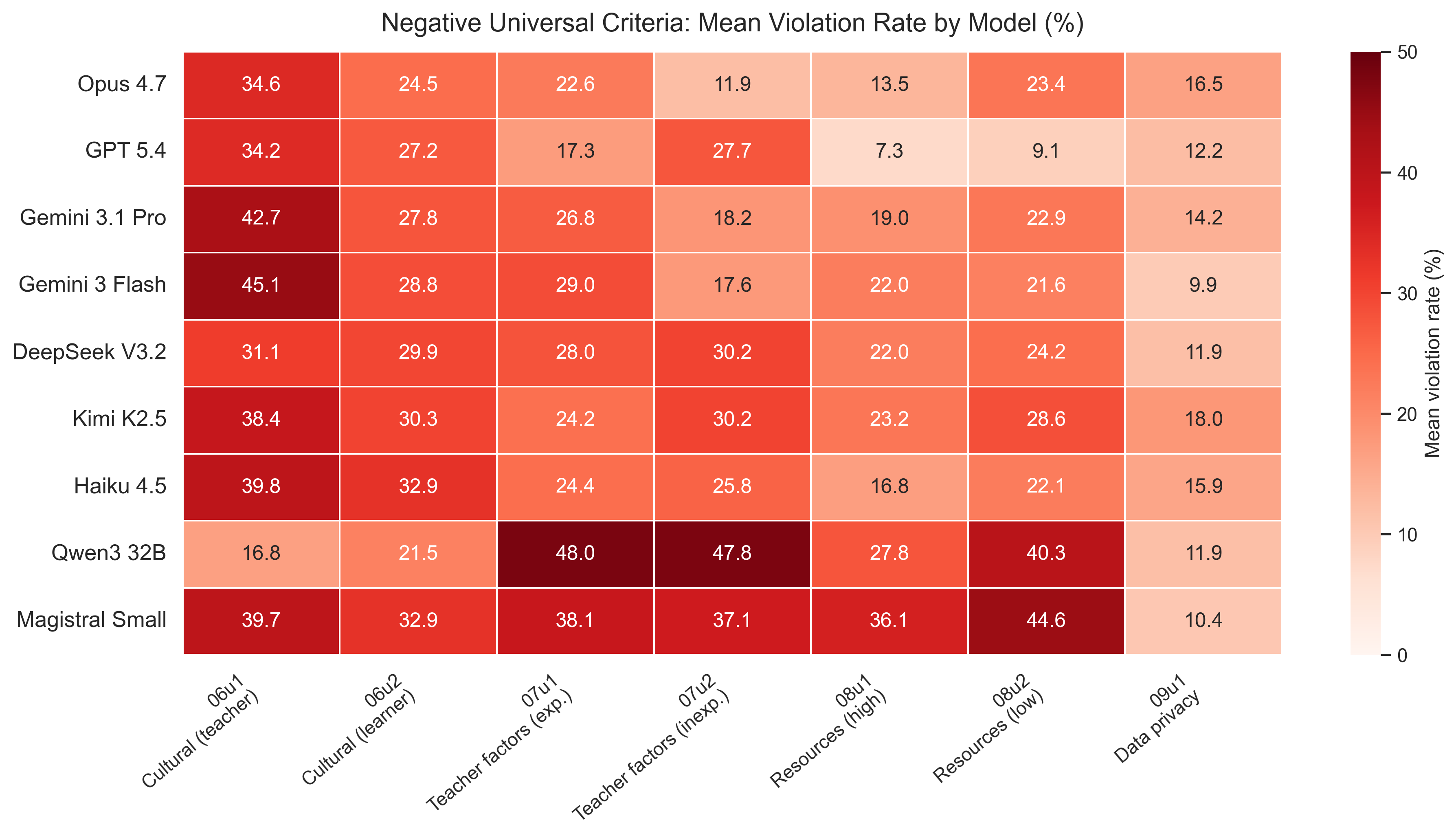}
    \caption{Mean violation rates of negative universal criteria by model. Cultural sensitivity (06u1/06u2), teacher factor (07u1/07u2), and resource appropriateness violations (08u1/08u2). All models violated universal criteria (appropriacy, CEFR level, cultural sensitivity, resource awareness, data privacy).}
    \label{fig:model-weaknesses}
\end{figure*}

\subsection{Scoring pipeline and judge validation}
\label{subsec:scoring}

Automated scoring is essential for benchmark reproducibility and scalability. We describe the scoring pipeline, judge optimization, and validation results below (full methodology in Appendix~H).

\subsubsection{Scoring pipeline} Each task response is scored on a per-criterion basis using an LLM-as-a-Judge pipeline. The judge receives the task input, metadata, rubric criterion, the model response under evaluation, and the expert reference answer. It returns a binary Pass/Fail verdict per criterion; the task-level score is the weighted sum of passed criteria divided by the sum of positive weights (negative criteria act as penalties; see Appendix~I). Temperature is set to zero for reproducibility \cite{dev2026simpler}.

\subsubsection{Judge selection} To select the production judge, we conducted two small experiments (see Appendix~H for details). Firstly, we used a stratified sample of 48 tasks (4 per competency) for which practitioner criterion-level scores were available from the validation study (the rater baseline---$N = 145$ practitioners contributed 1,023 criterion-level ratings across 443 items). We then tested three judge foundation models across four prompt variants crossing two dimensions: reference guidance (supplied vs.\ omitted) and prompt scaffolding mode (plain classifier vs.\ chain-of-thought + few-shot examples ("CoT")). This yielded 15 configurations to measure our candidate judge's performance: 9 for a Part A test (judge vs.\ rater majority vote on solver responses) and 6 for a Part B test (judge vs.\ expected verdicts on reference answers). Secondly, we also conducted a small stability analysis (3-resample test across 16 model $\times$ prompt configurations) which confirmed that classifier prompts consistently outperform "CoT" variants on verdict determinism \cite{dev2026simpler}. Based on convergent evidence across accuracy, stability, and cost, we selected Claude Sonnet 4.6 with the reference-guided classifier prompt as the production judge (see Appendix~H for details). To address the risk of any same-family (Claude) self-preference in our production LLM-as-a-judge setup, we note two design mitigations: (1) judging is not a holistic quality comparison but a reference-guided, per-criterion binary grading against a concrete expert answer, which constrains the room for stylistic preference; and (2) our judge selection was informed by both accuracy and stability across multiple model families, where our experiments found that cross-family agreement was comparable to within-judge resampling stability. We leave a more complete multi-model, multi-prompt self-preference audit for future work (see Appendix~A).

\subsubsection{Judge validation results} The production judge achieved Cohen's $\kappa = 0.746$ against human rater majority vote (67\%)---categorized as ``substantial agreement'' \cite{landiskoch1977}---with accuracy of 92.3\%, F1 = 0.932, and recall of 0.948 across 326 criteria with matched practitioner validation data. Note that we report agreement at multiple human-consensus thresholds (50\%, 67\%, 75\%, 100\%) to avoid arbitrary cutoff effects (full results in Appendix~H).

These results must be interpreted against the human inter-rater reliability ceiling. On criterion-level binary judgments, practitioners achieved Krippendorff's $\alpha = 0.362$ [95\% CI: 0.343, 0.380] with 89.8\% raw agreement. Low inter-annotator agreement is expected and well-documented for high-inference educational constructs \cite{thomas2026modernizinggroundtruthshifts}. As Messick \shortcite{messick1995validity} argues, the validity of an evaluation rests not on perfect rater consensus but on the extent to which score interpretations correlate with the real-world phenomena they purport to measure. Under this framework, convergent human-judge agreement (Cohen's $\kappa = 0.746$) constitutes strong validity evidence: the automated scorer captures the same pedagogical quality signal that trained practitioners detect. We note that this $\kappa$ is computed against a \emph{denoised} human majority-vote target, whereas the $\alpha = 0.362$ reflects raw \emph{pairwise} inter-rater agreement; agreement against a consensus target is mechanically higher than pairwise agreement, so the two statistics measure different quantities and are not directly comparable.

Following Thomas \shortcite{thomas2026modernizinggroundtruthshifts}, we interpret rater disagreement not as noise to be eliminated but as informative signal about construct complexity. High-inference pedagogical judgments---determining whether feedback appropriately diagnoses a learner's error, or whether a lesson sequence promotes meaningful engagement---inherently resist unanimous categorization. That 221 practitioners from 45 countries achieved 89.8\% raw agreement on such judgments, despite systematic differences in severity and pedagogical background, signals construct clarity: practitioners converge on what constitutes competent performance even while expressing their verdicts at different calibration points (Thomas 2025). Our observer protocol---calibration training, time-gated stage progression, multi-factor quality exclusion, and transparent demographic reporting---mirrors the rigor of established human observation systems (e.g. BROMP \cite{ocumpaugh2015bromp}), adapted for online expert panels.

The LLM judge matched or exceeded human inter-rater consistency on 7 of 9 tested configurations. As an internal validity check, both humans and the judge scored expert reference answers against the task rubric: humans assigned only an 80.9\% pass rate to reference criteria---substantially below the expected near-100\%---while the judge achieved 89.9\% accuracy. This confirms that even ``gold standard'' answers exhibit genuine ambiguity under rubric scrutiny, reflecting inherent construct complexity rather than pipeline failure.


\section{Evaluation results}
\label{sec:results}

We evaluated an initial selection of nine models across sizes and families. We ranked models by mean overall score across 1,000 tasks (3 runs per task, weighted by criterion importance). We computed 95\% confidence intervals via standard error. Our main findings show that L2-Bench provides reliable and meaningful measurement of AIED for second language learning design capabilities. It also provides diagnostic signal of two other targets: model weaknesses, and model performance across different geographic, economic, resource, age, and role contexts. We summarize below.  

\subsubsection{Main findings} 

Model results are summarized in Table~\ref{tab:model_ranking}, with the leaderboard showing a clear tiering by model size. A top tier of Claude Opus 4.7 (85.5\%), GPT 5.4 (84.1\%), and Gemini 3.1 Pro (83.4\%) leads; a mid-tier cluster of Gemini 3 Flash, DeepSeek V3.2, Kimi K2.5, and Haiku 4.5 falls in the 79--81\% range; and Qwen3 32B (65.8\%) and Magistral Small (50.7\%) trail substantially. The 95\% confidence intervals are tight (approximately $\pm0.7$pp), reflecting the large task set, and although no adjacent models' CIs overlap except within the mid-tier cluster, we stress that these CIs capture sampling variability over the 1{,}000 tasks, not measurement uncertainty arising from a noisy construct (criterion-level human $\alpha = 0.362$) or an imperfect judge. It is possible that the 1.4pp gap separating the top three models sit within plausible systematic-error bands from these sources, and so future work is planned to propagate judge or criterion uncertainty into these intervals (see Appendix~A).

Beneath the surface, competency-level performance is not uniform across models; Kruskal-Wallis tests per model determined that all models show significant differences across competencies ($p < 0.001$; see Table~\ref{tab:anova_kw_models}). All models perform relatively well on structured output tasks (lesson planning, activity planning) but less well on open-ended competencies (conversational exchange, giving feedback, evaluating performance), as shown in Figure~\ref{fig:model-performance-by-competency}. Notably, no one frontier model dominates both structured and open-ended tasks. As a conversational exchange partner, GPT 5.4, which ranks second overall, is outperformed by our leading mid-size model (Gemini 3 Flash) and both small models (Kimi K2.5 and Haiku 4.5). Observed weaknesses in open-ended, interactional competencies must be caveated by our current single-turn scoping; now we have a stable baseline of simple interactions, future work will involve multi-turn dataset production to establish whether this reflects a genuine capability gap (see Appendix~A).

\begin{table}[h]
\centering
\caption{ANOVA and Kruskal--Wallis results across models, testing whether per-model performance differs across the 12 competencies ($N = 12$ competencies per model; all $p < 0.001$ for both tests).}
\label{tab:anova_kw_models}
\begin{tabular}{lccc}
\toprule
Model & $F$-stat & $H$-stat & $\eta^2$ \\
\midrule
Haiku 4.5        & 19.65 & 190.46 & 0.180 \\
Opus 4.7         &  8.82 & 100.35 & 0.089 \\
DeepSeek V3.2    & 13.41 & 139.79 & 0.130 \\
Gemini 3 Flash     & 12.56 & 139.50 & 0.123 \\
Gemini 3.1 Pro       & 10.96 & 112.90 & 0.109 \\
GPT 5.4          & 10.13 & 107.06 & 0.101 \\
Kimi K2.5        & 12.03 & 141.37 & 0.118 \\
Magistral Small  & 20.97 & 187.77 & 0.189 \\
Qwen3 32B        & 20.31 & 167.49 & 0.184 \\
\bottomrule
\end{tabular}
\end{table}

We also performed verbosity analysis (see Appendix~I.4 for details), which is increasingly important because token budgets are fast becoming a bottleneck for AI evaluations \cite{ghosh2026evalbottleneck}. AIED evaluations are especially prone to this pressure because of chronic underresourcedness. In one variant of our leaderboard ranking, length-adjusted scores caused GPT 5.4's overall rank to dip from 2nd to 4th place, below Gemini 3.1 Pro and Gemini 3 Flash, suggesting it may be more burdensome on adopter resources at the current time.

Rank stability analysis was performed across variants, including a hard-items-only variant (the 267 tasks the top three models all scored below 80\% on) and the verbosity variant (see Appendix~I for details). High Kendall's $\tau$ values ($>0.7$) indicate that overall rankings are stable across variants, and top models remain at the top regardless of how the leaderboard is computed; lower values would suggest that certain variants reveal different capability orderings.

\subsubsection{Model weaknesses} 

\begin{figure*}[h]
    \centering
    \includegraphics[width=1\linewidth]{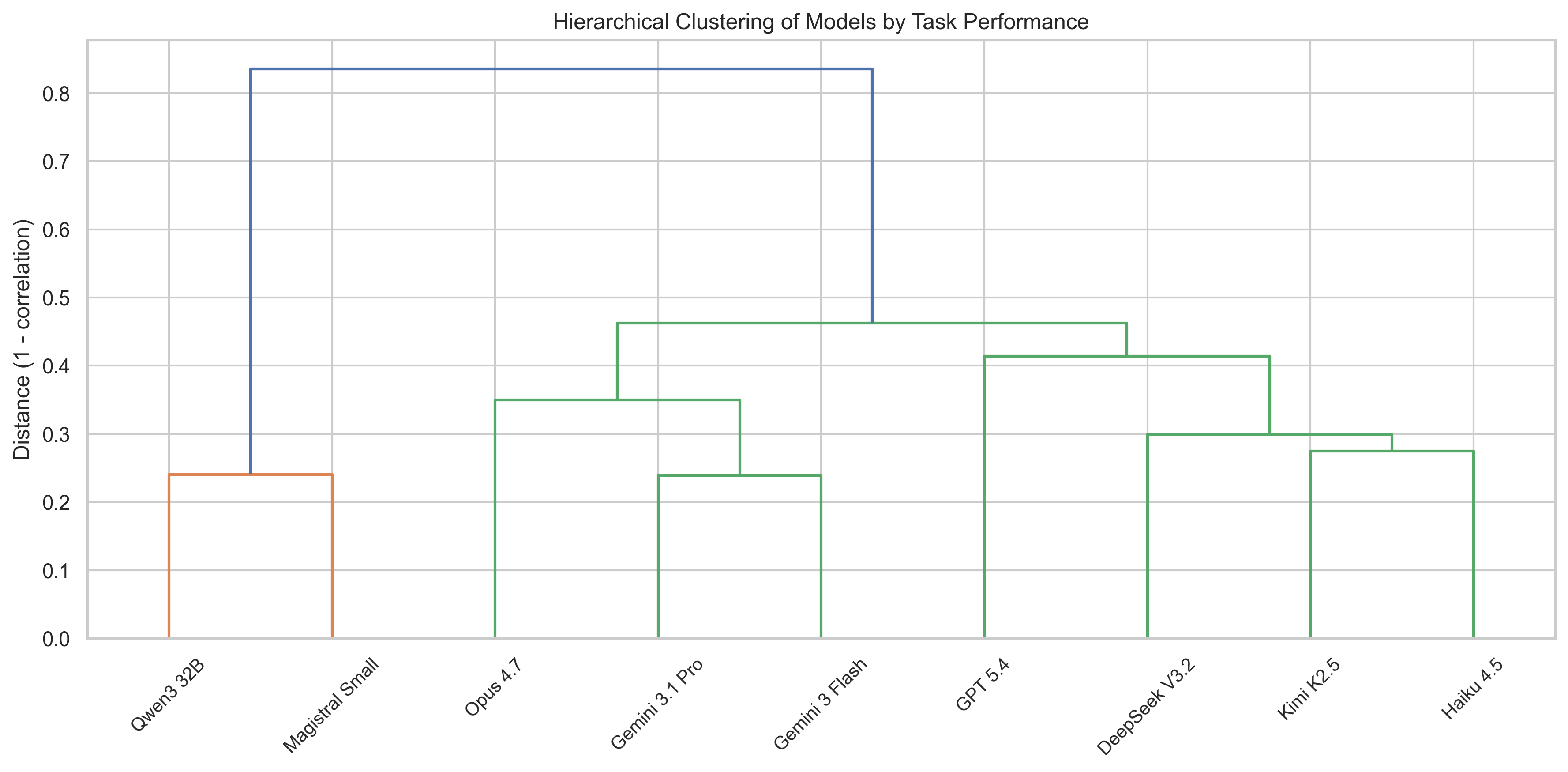}
    \caption{Hierarchical clustering of model response patterns. Gemini model families exhibit similar response patterns, while GPT 5.4 appears to respond to tasks in a manner more similar to smaller models (including DeepSeek V3.2) than its larger counterparts (Opus 4.7 and Gemini 3.1 Pro).}
    \label{fig:model-hierarchical-clustering}
\end{figure*}

Pairwise comparison of model performance confirms that models agree both on what is easy and what is hard. Patterns in model performance across the evaluation dataset yield clear, if perhaps surprising, hierarchical clustering (see Figure~\ref{fig:model-hierarchical-clustering}). On the one hand, Gemini model families appear to exhibit similar response patterns, although Gemini 3.1 Pro is generally more performant. On the other, GPT 5.4, which overall ranked second, appears to respond to tasks in a more similar manner to smaller models (including DeepSeek V3.2), and not its larger counterparts (Opus 4.7 and Gemini 3.1 Pro).

Most benchmarks are designed to tell us about successful performance \cite{eriksson2025trust}. However, as L2-Bench includes negative criteria (weight $< 0$), penalties are calculated, which provides signal about which criteria model outputs most often violate. Penalties are issued for, e.g., cultural insensitivity, privacy violations, or inappropriate resource assumptions. All models violated universal criteria (appropriacy, CEFR level, cultural sensitivity, resource awareness, data privacy) as shown in Figure~\ref{fig:model-weaknesses}. Understanding how and when models violate criteria surfaces valuable signal about what tasks may be the most challenging for AIED for language education and forms an important aspect of our ongoing work (see below). We acknowledge that, because negative criteria are relatively infrequent and enter only as penalties within a weighted-sum aggregate dominated by positive criteria, their impact is under-represented in the headline score; how best to surface safety-relevant violations in aggregate reporting is an open question we flag for future work (see Appendix~A).

\subsubsection{Performance Across Various Contexts}

We conducted analysis of model performance across various contexts to assess how task difficulty may be mediated by geographical, economic, and resourcedness indicators partially, fully, or not present in task prompts. Among top-tier models, we observe only modest overall performance differences. Unsurprisingly, smaller model performance varied more substantially as these models are less likely to be able to account for contextual factors as well as larger models.

Models appear to struggle more with pedagogically demanding low-proficiency contexts and in low-resource contexts, which present additional constraints (see Table~\ref{tab:resource_levels}). If models score lower on low-resource tasks, it suggests they default to resource-rich assumptions and struggle to adapt. This means L2-Bench provides signal about whether models can serve diverse educational contexts equitably.

\begin{table}[H]
\centering
\scriptsize
\caption{Model performance by combined resource level. Combined resource classification aggregates economic context, materials, internet, and devices into Low/Mixed/High.}
\label{tab:resource_levels}
\begin{tabular}{lccc}
\toprule
Model & Low & Mixed & High \\
\midrule
Opus 4.7         & 85.7 & 88.1 & 86.6 \\
GPT 5.4          & 86.9 & 88.1 & 85.7 \\
Gemini 3.1 Pro       & 81.3 & 83.3 & 84.3 \\
Gemini 3 Flash     & 78.6 & 78.8 & 80.0 \\
DeepSeek V3.2    & 78.2 & 80.0 & 78.8 \\
Kimi K2.5        & 77.1 & 80.8 & 78.7 \\
Haiku 4.5        & 76.8 & 80.5 & 78.5 \\
Qwen3 32B        & 50.9 & 56.3 & 58.1 \\
Magistral Small  & 35.0 & 39.9 & 40.2 \\
\bottomrule
\end{tabular}

\vspace{1em}

\scriptsize
\captionof{table}{Model performance by age group (Prim.\ = Primary, L-Sec.\ = Lower Secondary, U-Sec.\ = Upper Secondary, Tert.\ = Tertiary, i.e.\ post-secondary/higher education).}
\label{tab:age_groups}
\begin{tabular}{lccccc}
\toprule
Model & Prim. & L-Sec. & U-Sec. & Tert. & Adult \\
\midrule
Opus 4.7        & 86.3 & 86.6 & 86.0 & 86.9 & 86.1 \\
GPT 5.4         & 85.2 & 84.1 & 85.1 & 86.1 & 85.0 \\
Gemini 3.1 Pro      & 84.6 & 83.5 & 83.6 & 84.4 & 84.0 \\
Gemini 3 Flash    & 82.3 & 80.9 & 80.9 & 81.6 & 81.0 \\
DeepSeek V3.2   & 81.3 & 78.7 & 80.2 & 81.8 & 81.1 \\
Kimi K2.5       & 78.9 & 77.5 & 80.1 & 81.7 & 80.2 \\
Haiku 4.5       & 80.1 & 77.1 & 78.6 & 81.4 & 80.5 \\
Qwen3 32B       & 62.7 & 61.2 & 65.6 & 69.0 & 66.6 \\
Magistral Small & 49.4 & 42.1 & 50.6 & 52.9 & 51.3 \\
\bottomrule
\end{tabular}
\end{table}

We also analyzed model performance across human factors including learner age, and stakeholder persona (teachers, learners, assessment/curriculum designers). Age group is pedagogically significant; younger learners require fundamentally different approaches (e.g., play-based learning for pre-primary vs. academic writing for tertiary). The universal criteria apply stronger age-appropriateness weighting for younger learners (weight 10 for pre-primary/primary vs. 2 for adult). Lower scores on younger age groups indicate models struggle with age-appropriate pedagogy (see Table~\ref{tab:age_groups}). 

Models also perform across different stakeholder personas. Teacher-role tasks (the majority of L2-Bench) require pedagogical planning and expertise; Opus 4.7 performed best on most of these tasks. Learner-role tasks require adaptive conversation and appropriate language modeling; Gemini 3.1 Pro performed best on Learner-role tasks. Assessment and curriculum design tasks test more specialised knowledge; here Opus 4.7 and GPT 5.4 took leading positions. Differences across roles reveal that certain models are stronger in particular professional functions (see Appendix~I Table~\ref{tab:role_performance} for role-level breakdown).


\section{Discussion}
\label{sec:discussion}

L2-Bench establishes an evaluative baseline for AIED for second language education. But it also advances a novel evaluation methodology that may be suitable for several domains within education. It demonstrates the need for caution against the unreflective use of statistical measures commonly used in evaluation benchmark design (Cohen’s $\kappa$, Cronbach’s $\alpha$), or dogmatic adherence to conventional thresholds \cite{thomas2026modernizinggroundtruthshifts}. In educational spaces, it is often more illuminating to apply mixed-effects models, specifically to intraclass correlation measurements, because, although raters may have different absolute ratings for individual items, there may also be strong agreement about the relative ranking of item importance—this was the case for L2-Bench. Improving methodological literacy may increasingly prove essential to advancing the field of AIED evaluations. As He et al \shortcite{he2026judgingjudgeshumanvalidation} argue, rater disagreement is not necessarily “noise” . 

The success of this methodology may also suggest that there is no one correct pedagogical manifold towards which to optimize AIED. L2-Bench shows that models almost always receive a pass mark on consensus criteria that represent surface-level pedagogical appropriacy (e.g. formatting, staying on topic). It is when deeper competencies are required, in context, that differentiation is required. It may be, then, that there are as many pedagogical manifolds as there are educational situations. The goal of instance-level AIED evaluation, therefore, should be to assess how well AI models can regularly recognize their situation and then sample correctly from among multiple candidate pedagogical manifolds. L2-Bench attempts to move technical AIED evaluation in this direction.


\section{Conclusion}
\label{sec:conclusion}

Education—and language education specifically—is a widely recognized human right \cite{udlr1996}. As AI adoption in education accelerates, the public capacity to assess AI model performance on educational tasks is more important than ever. To this end, we openly share L2-Bench, the first comprehensive evaluation benchmark to assess model capabilities across tasks that comprise quality learning experience design in second language education. L2-Bench is, to our knowledge, the most extensively validated AIED benchmark evaluation. More than 200 expert practitioners provided high quality evidence that approves our novel taxonomy, the dataset itself (1000+ task-response pairs), and our evaluation rubrics. L2-Bench reveals important areas for improvement for all models, despite overall strong frontier performance. L2-Bench provides education stakeholders better methods to make more informed decisions about real-world AIED adoption, use, and governance, while advancing the maturing science of AI evaluations for education.

\clearpage

\section*{Acknowledgements}

L2-Bench was a collaborative effort and there are a great many people for their assistance and advice on this project.
With thanks to Professor Elizabeth Wonnacott, Department of Education, University of Oxford, who provided advice on statistical tests and experimental design to support the validation analysis, and to Beatrice Segura Harvey, ELT specialist, for contributing to L2-Bench dataset creation.
Thank you to our wider team and colleagues at Oxford University Press, in particular to Megan Gericke, for her support in facilitating the validation study and helping project delivery, and Dorian McCree for his continued support and advice since project inception.
We would like to thank all 221 education practitioners who contributed their time and expertise to the 2026 “OUP Global Practitioner Challenge”, in particular to those who ultimately helped to guide the development of L2-Bench and future iterations: Anna Król, Warsaw University of Technology; Yolanda Xavier, NOVA University Lisbon; Ozlem Terzioglu, British Council; Rachel Toncelli, Northeastern University; Wiktoria Allan, Berlin School of Economics and Law; Gülbahar Vidinel, Darüşşafaka Educational Institutions; Richard Delme Phillips; Deise Amaral, UFRGS; Serhii Andrusienko, Swiss NeuroLanguage Academy ANDRUSENKO.PRO; Hsin Yun Ho, University of Birmingham; Taiki Shimosakai, University of Birmingham; Rosangela Misciagna, Ca' Foscari University Venice; Alan S. Mackenzie, Mackenzie Education; David E. Ponce, Instituto de Lenguas Extranjeras; Adriana Maria Butnariu, Institut Camps Blancs; Marli Silva Pereira; Agnieszka Tyszkiewicz-Zora, University of Łódź; Carla Marmorale; Sabrina Boem; Monica Lunardon; Silvia Barlassina, IIS "Altiero Spinelli" Sesto San Giovanni; Masooma Amjad, University of Azad Jammu and Kashmir; Anna Maria Ruccolo, Milan ILC; John Bletsas, Hamlet EFL school; Tory S. Thorkelson, Sejong University; Magdalena Ecaterina Tolea, Ienachita Vacarescu National College ; Valentina Turrini, Lower secondary school "F.Cipriani" Nogara; Liliia Okhotina Okhotina; Magdalena Muszyńska; Jemma Hillyer, Oxford University Press; Ruyang Chloe Ye, Oxford University Press; Jasmin Anderson, Oxford University Press; Megan Hurley, Oxford University Press; Phil Davis, Oxford University Press.
Finally, we would like to thank all 39 postgraduate participants and organisers of the 2025 University of Birmingham “PGT SHAPE AI Challenge” for their hard work and valuable contributions in helping us iterate on early versions of L2-Bench, in particular the winners and runner-ups of the challenge: Venkata Vyjayanthi Pedapati (Vy), Yernur Niyetkaliyev, Aparajitha Magnesh, Manh Nguyen (Leo), Niamh Evans, Hsin-Yun Ho (Sydney), Sofía Muñoz, Saniya Saheer, Taiki Shimosakai, Yang Yu, and Dr. Liam Knight for his help facilitating this pilot.

\section*{Impact, Ethics, and Generative AI Statement}

\subsection*{IS1. Human Subjects Research for Data Validation} 

Research ethics were governed by Oxford University Press at all stages. A supplementary Institutional Review Board (IRB) from the Oxford Internet Institute (University of Oxford) departmental research ethics committee was sought (but ruled exempt) for the practitioner validation study. See Appendix~G.5 for further details describing: institutional oversight, UK GDPR compliance, informed consent and incentives; and Appendix~G.4 detailing: study instructions, calibration materials, and the time-gated workflow.

\subsection*{IS2. Broader Impacts} 

We envisage two real-world impacts of L2-Bench. Firstly, we anticipate that this work will contribute positively to the AIED ecosystem by providing a practitioner-validated, open-source benchmark and methodology that enables more rigorous assessment of AI capabilities in educational contexts, enabling education stakeholders to make more informed decisions about AIED adoption, use, and governance, and by extension promoting a more intentional AIED ecosystem predicated on effectiveness. Secondly, we hope that by open-sourcing our methods and benchmark, and openly reporting our positive and negative findings, we allow for greater scientific transparency in pursuit of a common goal—designing AIED that enables researchers and developers to respond to specific contexts of instruction and learning.

We remain attentive to several concerns. First, benchmarks assessing AI capabilities in education could, in principle, be repurposed to evaluate human educators; we note explicitly that our work focuses solely on AI system assessment and we have no products or interests in teacher evaluation. Second, L2-Bench is currently scoped to English as a target language and is built on frameworks of European origin, which may embed cultural and pedagogical assumptions that do not necessarily transfer to other target languages or to non-European traditions. Third, leaderboard rankings could inadvertently incentivise benchmark-specific optimisation rather than genuine pedagogical improvement. 

Unanticipated consequences may arise from applications of our competency taxonomy or evaluation methodology in ways we have not foreseen. We encourage researchers building on this work to consider the potential for dual-use applications, to examine their own assumptions about pedagogical quality, and to implement appropriate safeguards when deploying evaluation frameworks in educational contexts.

\subsection*{IS3. Generative AI Statement}

Generative AI tools were used in several aspects of this research. LLM usage methodology is fully disclosed, with Section 3 and Appendix~F.1 describing Claude models for hybrid human-AI task generation. All other uses limited to editing, table and figure finessing, bibliographic entry reformatting, and other matters of latex formatting.

For research support, GenAI assisted with: reviewing experimental design and identifying methodological improvements; research on statistical methods and their implementation; reviewing data processing pipelines and analysis iterations; and iterating on data visualisations. 

For manuscript preparation, GenAI assisted with: LaTeX table and equation formatting, grammar and spelling review, bibliographic entry reformatting, and website development that resulted in figure creation.

All substantive research decisions, interpretations, and conclusions remain solely the responsibility of the authors.



\bibliography{aaai2026}


\onecolumn
\appendix

\noindent\rule{\textwidth}{0.4pt}
\begin{center}
\textbf{\Large Appendix}
\end{center}
\vspace{0.5em}
\small
\begin{tabular}{@{}p{0.6cm}p{12.5cm}@{}}
\textbf{A} & \textbf{Limitations and Future Work} \\[0.3em]
\textbf{B} & \textbf{Code and Data} \\[0.3em]
\textbf{C} & \textbf{Glossary of Terms} \\[0.3em]
\textbf{D} & \textbf{Construct Development} \\
  & \hspace{1em}D.1~~Development of the L2-Bench Competency Taxonomy\\
  & \hspace{1em}D.2~~Initial Benchmarking Beyond Education \\
  & \hspace{1em}D.3~~Review of Existing Pedagogical and Professional Frameworks \\
  & \hspace{1em}D.4~~Selection and Structuring of Core Competencies \\
  & \hspace{1em}D.5~~Task-Led Development of Sub-Competencies and Criteria \\
  & \hspace{1em}D.6~~Emergence of Multiple Types of Evaluation Criteria \\
  & \hspace{1em}D.7~~Iteration and Refinement \\
  & \hspace{1em}D.8~~Development of Context Factor Model \\
  & \hspace{1em}D.9~~Transferability to Other Areas of Education \\
  & \hspace{1em}D.10~~Transferability to Other Languages \\[0.3em]
\textbf{E} & \textbf{L2-Bench Construct} \\
  & \hspace{1em}E.1~~Competencies, Sub-competencies, and Consensus Criteria \\
  & \hspace{1em}E.2~~Context Factors \\
  & \hspace{1em}E.3~~Universal Criteria \\[0.3em]
\textbf{F} & \textbf{L2-Bench Task Items} \\
  & \hspace{1em}F.1~~Item Production Process \\
  & \hspace{1em}F.2~~Task Criteria Design \\
  & \hspace{1em}F.3~~Task Examples \\
  & \hspace{1em}F.4~~Dataset Distribution \\[0.3em]
\textbf{G} & \textbf{Practitioner Validation Study} \\
  & \hspace{1em}G.1~~Overview and Research Objectives \\
  & \hspace{1em}G.2~~Participant Recruitment and Allocation \\
  & \hspace{1em}G.3~~Dataset Preparation and Item Allocation \\
  & \hspace{1em}G.4~~Study Platform and Procedure \\
  & \hspace{1em}G.5~~Study Ethics \\
  & \hspace{1em}G.6~~Statistical Methods \\
  & \hspace{1em}G.7~~Construct Validation Results \\
  & \hspace{1em}G.8~~Answer Preference Results \\[0.3em]
\textbf{H} & \textbf{Judge Building} \\
  & \hspace{1em}H.1~~Judge Prompt Design \\
  & \hspace{1em}H.2~~Judge Performance Experiment \\
  & \hspace{1em}H.3~~Judge Stability Experiment \\
  & \hspace{1em}H.4~~Judge Selection \\[0.3em]
\textbf{I} & \textbf{L2-Bench Results} \\
  & \hspace{1em}I.1~~Scoring Formula \\
  & \hspace{1em}I.2~~Statistical Methods \\
  & \hspace{1em}I.3~~Model Selection \\
  & \hspace{1em}I.4~~Score Variants \\
  & \hspace{1em}I.5~~Performance Across Task Contexts \\[0.3em]
\end{tabular}
\normalsize
\noindent\rule{\textwidth}{0.4pt}
\vspace{1.5em}

\clearpage

\section{Limitations and Future Work}
\label{app:limitations}

This appendix consolidates the limitations noted throughout the paper and the future work they motivate.

\paragraph{Single-turn.} L2-Bench items are single-turn task-response pairs. This design isolates a model's ability to produce a high-quality artefact given a fully specified context, but it cannot directly observe interactional competencies that only emerge over a dialogue---uptake of learner contributions, real-time repair, and interactional scaffolding. Single-turn scores may therefore under- or over-estimate a model's classroom-relevant interactional ability, and the open-ended/conversational weaknesses reported in the main findings should be read with this caveat. Therefore, an extension with multi-turn interactional tasks, including audio modalities common in language-learning design, will be addressed in future work.

\paragraph{Aggregate scoring and negative criteria.} An L2-Bench task score is the weighted sum of passed positive criteria, with negative criteria acting as penalties. Because most tasks contain few negative criteria and models rarely trigger them, the influence of negative criteria (which typically cover safety/appropriateness and therefore reveal the ways models are making mistakes) on the headline aggregate is small. A model can therefore score highly on L2-Bench while occasionally violating a low-frequency negative criterion. To better highlight these negative pedagogical instances then, we will pursue educational safeguarding / responsible AI benchmarking in future work.

\paragraph{Judge analysis, uncertainty and thresholds.} The production judge (Claude Sonnet 4.6) shares a model family with the top-ranked entrant (Claude Opus 4.7), raising a potential self-preference confound. While we discuss our mitigations in subsection~Judge Selection, we are actively pursuing a more complete judge analysis for future work, including conducting a more extensive multi-family, multi-prompt judge self-preference audit. Furthermore, although we report bootstrap confidence intervals on aggregate model scores (Table~\ref{tab:model_ranking}), we do not propagate judge- or criterion-level uncertainty into those intervals. Our future judge work will therefore quantify possible systematic biases introduced by the production judge model, and attempt to propagate any judge/criterion uncertainty into reported intervals. More broadly, this work will seek to establish more principled, domain-specific adequacy thresholds for AIED validation instruments, rather than relying on pre-registered gates.

\paragraph{English and European frameworks.} L2-Bench covers English as the target language and is grounded in frameworks of European origin. Extension to additional target languages and non-European pedagogical traditions is a natural area for future work, and is discussed at length in Appendix~D.

\paragraph{Model leaderboard.} At the time of initial benchmarking, the L2-Bench leaderboard reflects only nine frontier, mid-sized, and small models configured (see Appendix~I.3 for selection details). Expansion to evaluate a more diverse collection of models and model families is being actively pursued as future work (including open-sourcing scoring pipelines and datasets, see Appendix~B), with a leaderboard to be actively maintained at https://benchmarks.elt.edu.oup.com/.

\clearpage

\section{Code and Data}
\label{app:code}

L2-Bench is released as an open evaluation framework to support reproducibility and community extension. We open-source the full benchmark: the construct (competency taxonomy, and systematic rubrics), the task item dataset (1,000 task pairs, full rubrics and reference answers), the production LLM-as-a-Judge evaluation pipeline (including judge prompts used in production), the scored task-response pairs.

\paragraph{Repositories.} The evaluation codebase and benchmark dataset (tasks, rubrics, and reference answers) is available at \url{https://huggingface.co/datasets/OUP/l2-bench}.

\paragraph{Licensing.} The dataset, rubrics, and reference answers are released under the Creative Commons Attribution--ShareAlike 4.0 International licence (CC-BY-SA-4.0); the evaluation code is released under the MIT licence.

\paragraph{Maintenance and contamination countermeasures.} To mitigate pre-training contamination after open release, the dataset card publishes a canary GUID string that data collectors can exclude from training corpora, and items are constructed from novel combinations of context factors (Appendix~E) rather than reused public prompts. We commit to versioned releases with an identitically distributed held-out subset of items retained for future contamination audits, and will maintain the benchmark on a regular update cadence (see Appendix~A).

\clearpage

\section{Glossary of Terms}
\label{app:glossary}

\begin{table}[h]
\centering
\caption{Key terminology used in second language education and L2-Bench.}
\label{tab:glossary}
\small
\begin{tabular}{p{3cm}p{10cm}}
\toprule
\textbf{Term} & \textbf{Explanation} \\
\midrule
L1, L2 & L1 = first language or mother tongue; L2 = any additional language learned after the L1. \\
EAL, ESL & EAL = English as an Additional Language; ESL = English as a Second Language. Both terms are most commonly used when the learner resides in a country where English is the dominant language (e.g., the US or UK) but has a different L1. \\
EFL & EFL = English as a Foreign Language. Used when the learner is studying English in a context where it is not the dominant language (e.g., Spain or China). \\
ELT & ELT = English Language Teaching. Refers to the profession of teaching English and encompasses both EAL and EFL contexts. \\
Learning experience designer & In L2-Bench, this term encompasses roles that intentionally design the conditions shaping how people learn, including classroom and online teachers, materials developers (content or assessment creators), learning designers, and teacher trainers. \\
Language acquisition, learning, teaching & Language acquisition refers to the largely unconscious process of learning a language through immersion. Language learning implies a more conscious effort, often supported by a teacher. Language teaching is the purposeful activity of helping a learner acquire the language. \\
Competencies & A combination of knowledge, skills, and attitudes required to perform a role or occupational function successfully. In L2-Bench, competencies refer to those required by teachers and other language education practitioners. \\
Communicative competence & An individual's ability to convey meaning effectively across contexts, encompassing grammatical, sociolinguistic, discourse, and strategic competence. \\
CEFR & The Common European Framework of Reference for Languages, developed by the Council of Europe and widely recognized as an international standard for describing language proficiency (levels A1--C2). \\
Target Language & The language the learner aims to learn. In L2-Bench, both the Target Language and Language of Instruction are English. \\
\bottomrule
\end{tabular}
\end{table}

\clearpage

\section{Construct Development}
\label{app:construct-creation}

This appendix documents the development process for the L2-Bench competency framework and evaluation methodology.

\subsection{Development of the L2-Bench Competency Taxonomy}

The L2-Bench competency framework was developed through a multistage, iterative process combining insights from AI benchmarking in other domains, established pedagogical frameworks, and task-based analysis of authentic professional practice in language education.

\subsection{Initial Benchmarking Beyond Education}

Before engaging with education-specific competency frameworks, the team examined AI benchmarks in other applied, high-stakes domains, particularly health and medicine. These benchmarks offered early methodological guidance on how complex professional practice can be operationalized for AI evaluation. In particular, they highlighted the limitations of evaluations centered on factual knowledge retrieval or narrow task accuracy, and instead foregrounded the importance of assessing applied judgment, decision-making, and domain-specific reasoning. This cross-domain work shaped the early assumption that an educational benchmark would need to move beyond declarative knowledge toward performance in realistic, practice-based scenarios.

\subsection{Review of Existing Pedagogical and Professional Frameworks}

The team conducted a structured review of widely recognized competency and professional development frameworks in language education (Table~\ref{tab:frameworks-review}). This review was comparative rather than adoptive. The goal was to identify competencies and practices that recurred across frameworks, particularly those that cannot be reduced to knowledge easily retrieved from reference materials, since such tasks were expected to be weak discriminators in an AI benchmark.

This phase contributed to the decision to frame the benchmark around a broad \emph{learning experience designer} role, encompassing classroom teaching, materials development, assessment creation, feedback, learner support, and professional development.

\begin{table}[h]
\centering
\caption{Existing pedagogical and professional frameworks reviewed during L2-Bench development.}
\label{tab:frameworks-review}
\small
\begin{tabular}{p{5.5cm}p{7cm}}
\toprule
\textbf{Framework} & \textbf{Purpose} \\
\midrule
British Council CPD Framework for Teachers & Teacher professional development \\
Eaquals Framework for Language Teacher Training \& Development & Accreditation and teacher development \\
Cambridge English Teaching Framework (CETF) & Teacher development and assessment \\
European Profiling Grid / CEFR-aligned frameworks & Profiling qualifications and experience \\
Millin Competency Framework for Language Learning Materials Writing \cite{millin2023competency} & Materials writer development \\
\bottomrule
\end{tabular}
\end{table}

\subsection{Selection and Structuring of Core Competencies}

From this review, the team shortlisted a set of competencies intended to capture the main dimensions of professional practice shaping learners' experiences in second language education. These competencies were framed as \emph{what practitioners do} rather than \emph{what practitioners know}, and were designed to apply across multiple roles (e.g., teachers, materials writers, teacher trainers).

Each competency was then articulated into more granular sub-competencies, describing specific capabilities that could plausibly be assessed through observable task performance.

\subsection{Task-Led Development of Sub-Competencies and Criteria}

A defining feature of the framework's development was its \emph{task-led methodology}. Rather than finalizing sub-competencies and evaluation criteria in abstraction, the team designed an initial set of benchmark tasks aligned to each competency. These tasks reflected authentic scenarios in language education practice and drew on: (i) internal domain knowledge of common professional tasks; (ii) recurring themes from practitioner feedback; and (iii) the development team's own experience in pedagogical design and evaluation.

As tasks were created, they acted as stress tests for the emerging framework. Specifying what constituted a competent response exposed ambiguities, omissions, and overgeneralizations in early sub-competency definitions, leading to repeated cycles of revision. As a result, sub-competencies and evaluation criteria evolved in response to the demands of concrete task design rather than top-down theoretical modeling alone.

Another important dimension was that this framework was designed for AI systems simulating teacher competencies. Existing teacher frameworks make assumptions about human capabilities that need to be spelled out more explicitly for AI systems. For example, existing teacher competency frameworks do not cover the skills needed to carry out a simulated practice conversation with a learner---e.g., appropriate turn-taking, responses, expressions of support or encouragement.

\subsection{Emergence of Multiple Types of Evaluation Criteria}

During this iterative process, the team developed a structure involving different types of evaluation criteria associated with tasks and sub-competencies. This structure reflected similar approaches taken by other LLM benchmarking projects: the use of universal, consensus, and task-specific evaluation criteria (see, for example, OpenAI's HealthBench \cite{arora2025healthbenchevaluatinglargelanguage}). The aim is to evaluate performance across different levels of abstraction, making that evaluation as systematic as possible. Instead of developing different evaluation criteria for each task, we developed evaluation criteria for each of the sub-competencies relevant to that task.

The phrase ``consensus criteria'' usually refers to consensus among experts \cite{CHANG2025318}. To some extent, all three levels of evaluation criteria (universal, consensus, and task-specific) were developed through expert consensus, and the term is used here specifically to describe the criteria that apply to any task requiring a specific sub-competency. For example, under the sub-competency ``Assign an evaluation to a learner's performance'' (such as a mark or grade), there are three consensus criteria: (a) the evaluation is accurate according to the criteria or mark key, (b) the evaluation fits the required level of detail, and (c) if there are no clear evaluation criteria, the evaluation is based first on how well the performance communicates the intended meaning, and then on salient aspects of form.

Universal evaluation criteria are applied to all tasks, and include criteria such as age-appropriateness, CEFR-level appropriateness, and compliance with data privacy guidelines. Although these universal criteria are relevant to all tasks, the value attached to them can vary according to context factors. For example, age-appropriateness is given more weight when learners are in primary school.

Task-specific evaluation criteria identify important aspects of performance not captured through universal or consensus criteria. For example, while the consensus criterion might refer to practicing the target language of the lesson, the task-specific evaluation identifies what that target language is.

Each evaluation criterion is assigned a value identifying its significance in the task. Where an evaluation criterion identifies an aspect of performance to be avoided (e.g., references to eating pork in a Muslim environment), the value is negative. Crucially, this criteria structure was not fully specified at the outset, but emerged progressively in response to the practical challenges of scoring diverse, open-ended tasks in a consistent and pedagogically meaningful way.

\subsection{Iteration and Refinement}

The competency framework, sub-competencies, and criteria were further refined through internal review and early validation work, including pilot studies examining task authenticity and criteria adequacy. Feedback from these stages informed subsequent revisions, reinforcing the iterative nature of the methodology.

\subsection{Development of Context Factor Model}

An important component that emerged from the pilot study (reported in \cite{edgell2026accuracyrobustevaluationmethodology}), was the need for a more systematic way of dealing with context. We examined models of teacher knowledge, derived from Mishra and Koehler's TPACK model \cite{doi:10.1111/j.1467-9620.2006.00684.x}, based on the earlier work on pedagogical and content knowledge by Shulman \cite{doi:10.3102/0013189X015002004}. The TPACK model analyzes teacher knowledge into three categories: Technical, Pedagogical, and Content. Surrounding this is the notion of ``Contextual Knowledge,'' often referred to as XK. Nguyen et al.\ \cite{NGUYEN2025100290}examines XK more extensively, particularly the relationship between XK and other parts of the TPACK model. While this has not been the focus of the L2-Bench project, it prompted us to take a more systematic approach to context factors.

In a similar process to the development of the competency framework, we drew on domain expertise, pilot study feedback, and iterations of applying factors to different tasks in the model. We identified five broad categories of context factor for language learning tasks: Learning Purpose, Learner Characteristics, Learning Context, Resources, and Teacher Factors. Within those, we identified 33 context factor sub-types---ranging from Learner L1 script, to Class Size, to Exam Focus, to Internet Access, to Teacher Proficiency Level. We distinguished between factors frequently relevant to tasks and those important only occasionally. For example, ``session length'' would be important in many tasks, whereas ``peer relationships'' would less often be a significant factor. For each context factor sub-type, we developed a range of values relevant to task design and assessment criteria. For example, for ``class size'' we set five values: 2--3; 4--10; 11--20; 21--30; 30+. The factors and values are all subject to review and revision as validation activities continue.

The context factor model is significant for task and evaluation criteria development. For task development, the model ensures systematic coverage of the different contexts within which language education takes place around the world. Tasks are deliberately varied in the level of context specification provided, reflecting the variability of context information in real-world tasks. Evaluation criteria reflect context factors too, particularly among the task-specific criteria.

\subsection{Transferability to Other Areas of Education}

A substantial proportion of the L2-Bench competency framework reflects domain-general pedagogical knowledge, rather than content knowledge specific to language education. Core competencies such as planning, sequencing learning activities, scaffolding, providing feedback, and assessing progress are structurally similar across subjects. For example, a geography teacher sequencing map-reading skills from recognition to independent application engages in a learning design process structurally similar to an ELT teacher sequencing grammatical knowledge from presentation to guided practice to production.

However, sub-competencies and evaluation criteria become increasingly domain-specific as granularity increases. While high-level teaching practices may transfer across subjects, others are closely tied to domain-specific knowledge and theoretical frameworks. In language education, this specificity is reinforced by the theoretical infrastructure of second language acquisition (SLA), including constructs such as interlanguage or communicative competence. These concepts shape how language teachers design instruction, but they have no direct equivalents in many other subject domains. As a result, direct transplantation of ELT-specific competencies into other subjects would likely feel conceptually mismatched and pedagogically superficial.

For this reason, L2-Bench is best understood not as a universally transferable framework, but as an instantiation of a shared meta-framework. What is transferable is the structural logic of the framework: a hierarchical organization of competencies and sub-competencies, a task-based methodology grounded in authentic professional practice, and an iterative process for defining evaluation criteria. Extending L2-Bench to other areas of education would therefore require subject-specific communities to undertake the same process---identifying relevant competencies, refining sub-competencies, and developing evaluation criteria grounded in their own disciplinary theories and epistemologies.

\subsection{Transferability to Other Languages}

The L2-Bench competency framework draws primarily on standards and frameworks that originated in European contexts, including the CEFR, the British Council CPD Framework, and Eaquals. However, these frameworks are now widely used and adapted globally, and the CEFR in particular functions as an international reference point across many languages and educational systems. In this sense, L2-Bench reflects practices that have been stabilized through global uptake rather than norms confined to a single regional context.

That said, universality should not be assumed: while many higher-level pedagogical processes represented in the framework (such as planning, sequencing, assessment, and feedback) are broadly transferable, more fine-grained constructs may not generalize cleanly across languages or educational cultures. Differences in writing systems, orthographic depth, pragmatics, and discourse conventions mean that competencies developed primarily within English language teaching may require adaptation---or reconceptualization---when applied to other languages. Further investigation, ideally involving educators working across diverse linguistic and cultural contexts, would therefore be necessary to determine which elements of the framework generalize, which require modification, and which are fundamentally language-specific.

\clearpage

\section{L2-Bench Construct}
\label{app:construct}

This appendix provides the glossary, competency taxonomy, and context factor specification underlying L2-Bench.

\subsection{Competencies, Sub-competencies, and Consensus Criteria}

The L2-Bench competency taxonomy comprises 12 competencies, 31 sub-competencies, and 72 consensus criteria. For each competency we first list its sub-competencies, then give the consensus criteria (with weights) that any task tagged to each sub-competency inherits. Names are reproduced verbatim from the master taxonomy.

\subsubsection*{C01: Create a course plan for a learner or group of learners}
\noindent\textit{(3 sub-competencies, 9 consensus criteria)}

\noindent\textbf{Sub-competencies:}
\begin{itemize}\setlength\itemsep{0pt}
\item \textbf{01a:} Decide which learning goals are most important for the students' learning aims, their context, their needs and interests
\item \textbf{01b:} Organise learning goals into units and lessons
\item \textbf{01c:} Decide on learning experience design
\end{itemize}

\begin{table}[H]
\centering\small\setlength{\tabcolsep}{4pt}
\begin{tabular}{@{}lp{10.8cm}@{}}
\toprule
\textbf{Sub-competency} & \textbf{Consensus Criteria} \\
\midrule
\textbf{01a} & 01a-01 Includes reference to students' learning aim(s) - such an exam they're preparing for, a real world use of English (+7) \newline 01a-02 References the different areas of language learning i.e. grammar, vocabulary, speaking, listening, reading, writing, pronunciation, functional language, even if to say why they are not included (+7) \newline 01a-03 Selection of learning goals is appropriate for the time available, for level and the learning pace of the students (+7) \newline 01a-04 References information (if available) from tests, reports, learning analytics or other sources of data, which provide indications of relevant strengths and weaknesses (+4) \newline 01a-05 References information about students' interests - personal, academic or professional (+3) \newline 01a-06 Identifies when there is insufficient information about the context to create the course plan, and states the assumptions it is making about the context (+4) \\
\addlinespace
\textbf{01b} & 01b-01 learning goals progress in difficulty over the course, and complementary goals are grouped together into units and lessons (+4) \newline 01b-02 The plan includes the revisiting of learning goals from earlier units/lessons (+6) \\
\addlinespace
\textbf{01c} & 01c-01 Pace, recycling, assessment/progress checking, amount of practice, balance on online/in-person, use of digital tools, etc, are appropriate for the teaching context and learner profile (+8) \\
\bottomrule
\end{tabular}
\end{table}

\subsubsection*{C02: Plan a lesson}
\noindent\textit{(2 sub-competencies, 12 consensus criteria)}

\noindent\textbf{Sub-competencies:}
\begin{itemize}\setlength\itemsep{0pt}
\item \textbf{02a:} Decide on sequence and types of activities for the lesson, in order to create an effective learning experience
\item \textbf{02b:} Identify or create materials or other resources needed (including technical hardware and software)
\end{itemize}

\begin{table}[H]
\centering\small\setlength{\tabcolsep}{4pt}
\begin{tabular}{@{}lp{10.8cm}@{}}
\toprule
\textbf{Sub-competency} & \textbf{Consensus Criteria} \\
\midrule
\textbf{02a} & 02a-01 Includes an appropriate pattern in the choice and sequencing of activities, such as PPP (Presentation-Practice-Production), ESA (Engage, Study, Activate), TBLT (Pre-task, Task, Language focus, Repeat task) (+5) \newline 02a-02 The activities build up knowledge and skills relating to the learning goal (+6) \newline 02a-03 It has a clear structure and pedagogical approach appropriate for the student profile -- level, age, background (+8) \newline 02a-04 It provides instructions on how to set up and manage each of the activities (+9) \newline 02a-05 It balances a focus on form with a focus on function (+6) \newline 02a-06 It helps develop students' own learning skills through metacognitive reflection, self-assessment or other techniques (+5) \newline 02a-07 There is a sequence of activities with realistic timings (+6) \newline 02a-08 It supports students towards improved proficiency in the way it introduces language, moves from more scaffolded to less scaffolded activities (+7) \newline 02a-09 It provides activities that are likely to engage students of this profile, because of the themes and activity design, such as level of difficulty and interactivity (+8) \newline 02a-10 Identifies when there is insufficient information about the context to create a lesson plan, and states the assumptions it is making about the context (+4) \\
\addlinespace
\textbf{02b} & 02b-01 The resources are an appropriate length for the time and level of the activity (+7) \newline 02b-02 The texts are realistic within the constraints of the level and context (+7) \\
\bottomrule
\end{tabular}
\end{table}

\subsubsection*{C03: Plan an activity}
\noindent\textit{(6 sub-competencies, 9 consensus criteria)}

\noindent\textbf{Sub-competencies:}
\begin{itemize}\setlength\itemsep{0pt}
\item \textbf{03a:} Decide on most suitable type of activity
\item \textbf{03b:} Provide appropriate level of scaffolding
\item \textbf{03c:} Identify or create materials or other resources needed (including technical hardware and software)
\item \textbf{03d:} Create a key for evaluating student responses
\item \textbf{03e:} Provide instructions on how to run the activity
\item \textbf{03f:} Integrate activity with other activities in the lesson
\end{itemize}

\begin{table}[H]
\centering\small\setlength{\tabcolsep}{4pt}
\begin{tabular}{@{}lp{10.8cm}@{}}
\toprule
\textbf{Sub-competency} & \textbf{Consensus Criteria} \\
\midrule
\textbf{03a} & 03a-01 Activity suits the learning goal, the stage in the lesson, the type of lesson and the profile of the learners (+8) \\
\addlinespace
\textbf{03b} & 03b-01 Activity includes scaffolding at earlier stages of learning (+6) \newline 03b-02 Scaffolding is reduced appropriately as the learners show a degree of mastery of the learning goal (+4) \\
\addlinespace
\textbf{03c} & 03c-01 The resources match the details of the request, and are appropriate for the level, the teaching context, the profile of the learners (+8) \newline 03c-02 Where there are a number of resources, they have a coherent theme, such as the same topic or scenario (+5) \\
\addlinespace
\textbf{03d} & 03d-01 Indicates what range of answers will be accepted, provides rubrics where appropriate, indicates how marks to be given if applicable (+7) \\
\addlinespace
\textbf{03e} & 03e-01 The instructions are clear and include an indication of likely timings of the activity (+9) \\
\addlinespace
\textbf{03f} & 03f-01 It makes use of key language (e.g. vocabulary or grammar) in previous activities (+6) \newline 03f-02 Explicitly references skills or language points covered in previous lessons where important for this activity (+3) \\
\bottomrule
\end{tabular}
\end{table}

\subsubsection*{C04: Manage activities within a class}
\noindent\textit{(2 sub-competencies, 2 consensus criteria)}

\noindent\textbf{Sub-competencies:}
\begin{itemize}\setlength\itemsep{0pt}
\item \textbf{04a:} Check that instructions for activities are understood and followed
\item \textbf{04b:} Organise learners into pairs, groups, assign roles
\end{itemize}

\begin{table}[H]
\centering\small\setlength{\tabcolsep}{4pt}
\begin{tabular}{@{}lp{10.8cm}@{}}
\toprule
\textbf{Sub-competency} & \textbf{Consensus Criteria} \\
\midrule
\textbf{04a} & 04a-01 Instructions are clear and appropriate for the level; there is some process for checking that the learners have understood and can access the instructions as they complete the activity (+7) \\
\addlinespace
\textbf{04b} & 04b-01 Learners are put into pairs or groups according to some principles that suit the task and class (e.g. similar levels together, or mixed levels, by their own choice, or by the teacher's choice) (+5) \\
\bottomrule
\end{tabular}
\end{table}

\subsubsection*{C05: Present language learning points}
\noindent\textit{(1 sub-competency, 8 consensus criteria)}

\noindent\textbf{Sub-competencies:}
\begin{itemize}\setlength\itemsep{0pt}
\item \textbf{05a:} Present language learning points effectively
\end{itemize}

\begin{table}[H]
\centering\small\setlength{\tabcolsep}{4pt}
\begin{tabular}{@{}lp{10.8cm}@{}}
\toprule
\textbf{Sub-competency} & \textbf{Consensus Criteria} \\
\midrule
\textbf{05a} & 05a-01 The language point is presented and explained clearly (+10) \newline 05a-02 There is an approach to teaching the language point which is appropriate for the context: inductive, deductive approach or blended approach (+4) \newline 05a-03 The presentation includes steps to activate learners' prior knowledge before introducing new items (+5) \newline 05a-04 Language items are introduced in a meaningful situation or text that gives the learner cues for meaning and use (+7) \newline 05a-05 Explanations include aspects of meaning, use and form of the language being presented (+8) \newline 05a-06 Visual aids (e.g. images, bold text, or text in different colours) are used to make the meaning, form, and use of language clearer when necessary (+5) \newline 05a-07 The instructions include suggestions to check learners' understanding of the meaning, form, and use of the language being presented (+5) \newline 05a-08 If the approach is an inductive one, the questions and prompts help the learner to work out the learning point from the examples given (+8) \\
\bottomrule
\end{tabular}
\end{table}

\subsubsection*{C06: Act as a conversational exchange partner (spoken or written)}
\noindent\textit{(2 sub-competencies, 5 consensus criteria)}

\noindent\textbf{Sub-competencies:}
\begin{itemize}\setlength\itemsep{0pt}
\item \textbf{06a:} Respond appropriately for the role and context
\item \textbf{06b:} Identify when the learner is struggling and respond appropriately
\end{itemize}

\begin{table}[H]
\centering\small\setlength{\tabcolsep}{4pt}
\begin{tabular}{@{}lp{10.8cm}@{}}
\toprule
\textbf{Sub-competency} & \textbf{Consensus Criteria} \\
\midrule
\textbf{06a} & 06a-01 Makes use of key language where appropriate (+6) \newline 06a-02 Responds in real time (not too delayed) (+8) \newline 06a-03 Responds appropriately given information that the learner gave earlier in the conversation (+6) \newline 06a-04 Does not include teacher explanations or rationales for its response ($-$7) \\
\addlinespace
\textbf{06b} & 06b-01 Simplifies responses or uses L1 when learner doesn't respond appropriately (+8) \\
\bottomrule
\end{tabular}
\end{table}

\subsubsection*{C07: Evaluate a student's performance}
\noindent\textit{(1 sub-competency, 3 consensus criteria)}

\noindent\textbf{Sub-competencies:}
\begin{itemize}\setlength\itemsep{0pt}
\item \textbf{07a:} Assign an evaluation of the performance as required - from very general 'ok/not ok', a CEFR level, to detailed marks
\end{itemize}

\begin{table}[H]
\centering\small\setlength{\tabcolsep}{4pt}
\begin{tabular}{@{}lp{10.8cm}@{}}
\toprule
\textbf{Sub-competency} & \textbf{Consensus Criteria} \\
\midrule
\textbf{07a} & 07a-01 The evaluation is accurate according to the criteria or mark key (+10) \newline 07a-02 The evaluation fits the required level of detail (+6) \newline 07a-03 If there are no clear evaluation criteria, the evaluation is based first on how well the performance communicates the intended meaning, and then on salient aspects of form. (+4) \\
\bottomrule
\end{tabular}
\end{table}

\subsubsection*{C08: Give feedback}
\noindent\textit{(5 sub-competencies, 6 consensus criteria)}

\noindent\textbf{Sub-competencies:}
\begin{itemize}\setlength\itemsep{0pt}
\item \textbf{08a:} Identifies flaws and errors, and where possible diagnoses causes of error
\item \textbf{08b:} Prioritise areas that need feedback
\item \textbf{08c:} Provide explanations, models or hints to help learners improve
\item \textbf{08d:} Provide activities that help learners to improve their performance
\item \textbf{08e:} Include feedback on the positive aspects of the learner's performance
\end{itemize}

\begin{table}[H]
\centering\small\setlength{\tabcolsep}{4pt}
\begin{tabular}{@{}lp{10.8cm}@{}}
\toprule
\textbf{Sub-competency} & \textbf{Consensus Criteria} \\
\midrule
\textbf{08a} & 08a-01 Estimates the likely causes of the error - e.g. gaps in knowledge, or skill proficiency (+5) \\
\addlinespace
\textbf{08b} & 08b-01 Only provides feedback for those areas that are most important for the learning goal and the learner profile (+5) \newline 08b-02 Explains the basis for the prioritisation when this seems helpful (+4) \\
\addlinespace
\textbf{08c} & 08c-01 Feedback includes explanations, models or hints that the learner can use to improve (+8) \\
\addlinespace
\textbf{08d} & 08d-01 Points to or provides activities that learners can do to improve their performance (+7) \\
\addlinespace
\textbf{08e} & 08e-01 Includes feedback on at least one positive aspect of the learner's performance (+6) \\
\bottomrule
\end{tabular}
\end{table}

\subsubsection*{C09: Track progress}
\noindent\textit{(2 sub-competencies, 3 consensus criteria)}

\noindent\textbf{Sub-competencies:}
\begin{itemize}\setlength\itemsep{0pt}
\item \textbf{09a:} Collect data (including samples) of learning/performance over time
\item \textbf{09b:} Analyse patterns of progress for different learning goals
\end{itemize}

\begin{table}[H]
\centering\small\setlength{\tabcolsep}{4pt}
\begin{tabular}{@{}lp{10.8cm}@{}}
\toprule
\textbf{Sub-competency} & \textbf{Consensus Criteria} \\
\midrule
\textbf{09a} & 09a-01 There is data - scores and samples of performance - tagged for different learning goals (+8) \\
\addlinespace
\textbf{09b} & 09b-01 There are insights into learning progress, derived from the data (+10) \newline 09b-02 The analysis of the data is related to recognised or established standards (+5) \\
\bottomrule
\end{tabular}
\end{table}

\subsubsection*{C10: Manage the social-emotional aspects of the learners}
\noindent\textit{(2 sub-competencies, 7 consensus criteria)}

\noindent\textbf{Sub-competencies:}
\begin{itemize}\setlength\itemsep{0pt}
\item \textbf{10a:} Identify/diagnose the emotional status of the learner(s) -- happy, bored, confused, distracted/disengaged, etc
\item \textbf{10b:} Implement interventions to address any emotional issues
\end{itemize}

\begin{table}[H]
\centering\small\setlength{\tabcolsep}{4pt}
\begin{tabular}{@{}lp{10.8cm}@{}}
\toprule
\textbf{Sub-competency} & \textbf{Consensus Criteria} \\
\midrule
\textbf{10a} & 10a-01 Has a process for monitoring the emotional status of the learners (+7) \newline 10a-02 The interlocutor is able to identify learner's emotions (+5) \newline 10a-03 Provides information that helps the teacher to identify the emotional status of the learner (+2) \\
\addlinespace
\textbf{10b} & 10b-01 The interlocutor is able to show understanding and empathy (+7) \newline 10b-02 The interlocutor takes steps to raise learners' awareness of self-efficacy (+5) \newline 10b-03 The interlocutor takes steps to develop learners' understanding of self-regulated learning (+4) \newline 10b-04 Responds actively to emotional issues indicated by the monitoring - both short-term and long-term (+2) \\
\bottomrule
\end{tabular}
\end{table}

\subsubsection*{C11: Create assessments}
\noindent\textit{(3 sub-competencies, 6 consensus criteria)}

\noindent\textbf{Sub-competencies:}
\begin{itemize}\setlength\itemsep{0pt}
\item \textbf{11a:} Decide on learning goals to be assessed in each assessment
\item \textbf{11b:} Decide on the types of tasks, and their organisation
\item \textbf{11c:} Create a mark scheme
\end{itemize}

\begin{table}[H]
\centering\small\setlength{\tabcolsep}{4pt}
\begin{tabular}{@{}lp{10.8cm}@{}}
\toprule
\textbf{Sub-competency} & \textbf{Consensus Criteria} \\
\midrule
\textbf{11a} & 11a-01 Learning goals to be assessed are appropriate for the CEFR level, the learning aim, the teaching context and learner profile (+7) \newline 11a-02 Identifies when there is insufficient information about the context to create appropriate assessments, and states the assumptions it is making about the context (+4) \\
\addlinespace
\textbf{11b} & 11b-01 The types of tasks are appropriate for the learning goal (+10) \newline 11b-02 The sequence and organisation of tasks is practical for test administration within the context, including the duration of the test (+8) \\
\addlinespace
\textbf{11c} & 11c-01 The mark scheme indicates acceptable answers, rubrics where appropriate, marks available and how to award them, grade thresholds where applicable (+10) \newline 11c-02 The mark scheme includes rubrics for assessing spoken and written production if part of the test (+7) \\
\bottomrule
\end{tabular}
\end{table}

\subsubsection*{C12: Support professional development of the teacher}
\noindent\textit{(2 sub-competencies, 2 consensus criteria)}

\noindent\textbf{Sub-competencies:}
\begin{itemize}\setlength\itemsep{0pt}
\item \textbf{12a:} Evaluate a teacher's activity
\item \textbf{12b:} Provide advice and guidance on how to teach better or address an issue
\end{itemize}

\begin{table}[H]
\centering\small\setlength{\tabcolsep}{4pt}
\begin{tabular}{@{}lp{10.8cm}@{}}
\toprule
\textbf{Sub-competency} & \textbf{Consensus Criteria} \\
\midrule
\textbf{12a} & 12a-01 Includes reference to how effectively the students engaged in the activities, the choice of activities, classroom management, use of materials and resources, response to specific needs of students in the class, checking of learning, and different strategies used (+10) \\
\addlinespace
\textbf{12b} & 12b-01 Is appropriate for the professional experience and qualifications of the teacher (+10) \\
\bottomrule
\end{tabular}
\end{table}

\subsection{Context Factors}
\label{app:context-factors}

L2-Bench tasks are parameterized by 33 context factor variables across five dimensions. Table~\ref{tab:context-factors} lists each factor with its permitted values; factors marked \textit{Frequent} appear in a majority of tasks, while \textit{Less frequent} factors are sampled more sparingly.

\begin{table}[H]
\centering\small\setlength{\tabcolsep}{4pt}
\caption{Context factor taxonomy. Values are semicolon-separated; bracketed prompts (e.g.\ [specify exam]) are completed per task.}
\label{tab:context-factors}
\begin{tabular}{@{}llp{7.4cm}@{}}
\toprule
\textbf{Dimension} & \textbf{Factor} & \textbf{Values} \\
\midrule
\textbf{Learner Characteristics} & Age group & Pre-primary; Primary; Lower-Secondary; Upper-Secondary; Tertiary; Adult \\
 & Learner L1 script & Latin script; non-Latin script \\
 & Learner L1 literacy & Literate; No/Limited literacy \\
 & Learning needs & Special Education Needs; No Special Education Needs \\
 & Learner proficiency & A1; A2; B1; B2; C1; C2 \\
\addlinespace
\textbf{Learning Purpose} & Curriculum focus & General English; Exam preparation [specify exam, e.g. IELTS]; English for Specific Purposes [specify purpose, e.g. Engineering]; Content and Language Integrated Learning [specify subject, e.g. Science] \\
 & Skill focus & Integrated skills and systems; Specific skills and systems [specify skill and/or system, e.g. writing] \\
\addlinespace
\textbf{Learning Context} & Setting type & Mainstream education; University or FE; Language School [e.g. Private Language School]; self-guided learning \\
 & Class configuration & Individual; Group \\
 & Class L1 composition & Shared L1; Mixed L1s \\
 & Class size & 2-3;4-10; 11-20; 21-30; 30+ \\
 & Delivery mode & Face-to-face; Online; Blended \\
 & Learning scheduling & Synchronous; Asynchronous; Sync-Async \\
 & Contact hours-week & <2 hours; 2-5 hours, 5-10 hours, 10+ hours \\
 & Session length & <2 hours; 2-5 hours, 5-10 hours, 10+ hours \\
 & Physical environment & Dedicated classroom; Shared/flexible space \\
 & English-speaking community & Living in English-speaking country; English-speaking community nearby; No local English speakers \\
 & Peer relationships & Collaborative/supportive; Competitive; Neutral/disconnected \\
 & Cultural attitudes towards English & Highly valued; Neutral; Resistant/low motivation \\
 & L2 policy environment & English mandated; English encouraged; English contested/politicized \\
 & Institution/teacher culture & Front of class instruction; Facilitative/Collaborative teaching \\
 & Emergency/crisis status & Post-conflict; Refugee education \\
\addlinespace
\textbf{Teacher Factors} & Proficiency & A2;B1;B2;C1+ \\
 & Teacher L1 relationship & Teacher shares L1 with students; Teacher doesn't share L1 with students \\
 & Teaching experience & Experienced/well-trained; Inexperienced/training-deficient \\
 & Digital competence & High; Developing; Low \\
 & Curriculum autonomy & Full autonomy; Moderate autonomy; Prescribed curriculum \\
\addlinespace
\textbf{Resources} & Economic context & High resource area; Low resource area \\
 & Materials & Textbooks available; No textbooks \\
 & Projection screen & Available; Not available \\
 & Internet & Available in class; Available at home; Available in class and at home; Variable across class \\
 & Individual devices & Available in class; Available at home; Available in class and at home; Variable across class \\
 & Student digital competence & High; Developing; Low \\
\bottomrule
\end{tabular}
\end{table}

\subsection{Universal Criteria}

Nine universal criteria apply to all L2-Bench tasks with context-conditional weights, summarised in Table~\ref{tab:universal-criteria} (context-conditional variants can be found in Table~\ref{tab:universal-variants}). Criteria 06u--09u are negatively weighted because they penalize the \textit{presence} of undesirable features.

\begin{table}[H]
\centering
\caption{Universal Criteria summary.}
\label{tab:universal-criteria}
\small
\begin{tabular}{clcc}
\toprule
\textbf{ID} & \textbf{Criterion} & \textbf{Condition} & \textbf{Weight} \\
\midrule
01u & Response appropriate for learner age & Young learners & +10 \\
02u & Language appropriate for CEFR level & Learner-facing & +9 \\
03u & Response appropriate for learner characteristics & SEN & +10 \\
04u & Response appropriate for learning purpose & Specialized & +10 \\
05u & Response appropriate for learning context & Full context & +10 \\
06u & Response NOT appropriate for cultural sensitivities & Learner-facing & $-10$ \\
07u & Response NOT appropriate for teacher factors & Inexperienced & $-10$ \\
08u & Response NOT appropriate for available resources & Low resource & $-10$ \\
09u & Response does not comply with data privacy & Personal data & $-10$ \\
\bottomrule
\end{tabular}
\end{table}

Each universal criterion resolves to one of several \textit{conditional variants} depending on the task's context factors and the role(s) of the person the response is aimed at. The applicable variant, its triggering condition, and its weight are given verbatim in Table~\ref{tab:universal-variants}.

\begin{table}[t]
\centering\small\setlength{\tabcolsep}{4pt}
\caption{Universal criteria and their context-conditional variants. The trigger condition describes, in plain terms, when each variant's weight applies (see Table~\ref{tab:context-factors} for the underlying context factors).}
\label{tab:universal-variants}
\begin{tabular}{@{}lp{4.3cm}p{5.6cm}c@{}}
\toprule
\textbf{ID} & \textbf{Criterion} & \textbf{Trigger condition} & \textbf{Weight} \\
\midrule
\textbf{01u1} & The response is appropriate for the learner age. & Younger learners (pre-primary or primary) & +10 \\
\textbf{01u2} &  & Secondary-age learners & +5 \\
\textbf{01u3} &  & Adult or tertiary learners & +2 \\
\addlinespace
\textbf{02u1} & The language is appropriate for the CEFR level. & Response is aimed at the learner & +9 \\
\textbf{02u2} &  & Response is aimed at the teacher or another professional & +2 \\
\addlinespace
\textbf{03u1} & The response is appropriate for other learner characteristics (e.g.\ L1 script, L1 literacy, learning needs). & Learner has special educational needs & +10 \\
\textbf{03u2} &  & Learner has no special educational needs & +5 \\
\addlinespace
\textbf{04u1} & The response is appropriate for the learning purpose. & Specialized purpose (exam preparation, English for specific purposes, or CLIL) & +10 \\
\textbf{04u2} &  & General English & +5 \\
\addlinespace
\textbf{05u1} & The response is appropriate for the learning context (e.g.\ setting, class configuration, delivery mode, scheduling, educational culture). & Context is fully specified & +10 \\
\textbf{05u2} &  & Context is only partially specified or minimal & +5 \\
\addlinespace
\textbf{06u1} & The response is not appropriate for the cultural sensitivities (e.g.\ food, relationships, appearance, behaviour norms, holidays, geo-political context). & Response is aimed at the teacher or another professional & $-5$ \\
\textbf{06u2} &  & Response is aimed at the learner & $-10$ \\
\addlinespace
\textbf{07u1} & The response is not appropriate for the teacher factors (e.g.\ L1 relationship, experience and training). & Teacher is experienced and well-trained & $-5$ \\
\textbf{07u2} &  & Teacher is inexperienced or training-deficient & $-10$ \\
\addlinespace
\textbf{08u1} & The response is not appropriate for the available resources (materials, projection screen, internet, devices). & Resources are readily available (well-resourced setting) & $-5$ \\
\textbf{08u2} &  & Resources are scarce (low-resource setting) & $-10$ \\
\addlinespace
\textbf{09u1} & The response does not comply with data privacy guidelines. & Task involves handling personal data & $-10$ \\
\bottomrule
\end{tabular}
\end{table}

\clearpage

\section{L2-Bench Task Items}
\label{app:items}

This appendix details the item production process, task criteria design principles, and representative task examples.

\subsection{Item Production Process}
\label{app:item-production}

L2-Bench items (tasks, task criteria, and reference answers) are produced through a hybrid human-AI authoring approach modeled on publishing workflows:

\begin{enumerate}
    \item \textbf{Design:} Language pedagogy experts create hand-crafted task exemplars and establish prompt templates for task and reference answer generation.
    \item \textbf{Draft:} Foundation models with agent scaffolding generate candidate tasks, task criteria, and reference answers using the prompt templates and examples.
    \item \textbf{Review:} Experts iteratively refine generated content; modifications trigger regeneration cycles.
    \item \textbf{Approve:} A separate expert validates the reviewed items for pedagogical soundness.
    \item \textbf{Publish:} Items are published in version-controlled benchmark dataset releases.
\end{enumerate}

Items are produced in batches of no more than 144 (12 per competency) to enable iterative improvement of prompt templates and accumulation of approved examples.

To mitigate "criteria drift" during development (where grading outputs / creating reference answers helps authors refine criteria, but authors need evaluation criteria to grade outputs / create reference answers), we established consensus criteria and universal criteria \textit{before} generating task-specific criteria.

\subsection{Task Criteria Design}

Task criteria capture requirements specific to individual tasks that are not covered by consensus criteria (sub-competency-level) or universal criteria (domain-wide). Each task rubric therefore comprises three independent layers:

\begin{table}[h]
\centering
\small
\begin{tabular}{lll}
\toprule
\textbf{Layer} & \textbf{Scope} & \textbf{Example} \\
\midrule
Task criteria & Specific to this task & ``Creates a complete 45-min lesson plan'' (+10) \\
Consensus criteria & All tasks for this sub-competency & ``Activities build knowledge/skills for goal'' (+6) \\
Universal criteria & All tasks in L2-Bench & ``Language appropriate for CEFR level'' (+9) \\
\bottomrule
\end{tabular}
\end{table}

\textbf{Weighting Guidelines.} Criteria weights range from $-10$ to $+10$: $+10$ = essential; $+5$ = important but not central; $+2$ = nice to have; $-2$ to $-10$ = undesirable features (severity-scaled).

\textbf{Independence Constraint.} Task criteria must be independent of consensus and universal criteria to avoid double-counting.

\subsection{Task Examples}

Below are two representative L2-Bench tasks illustrating the three-layer criteria structure.

\paragraph{Example 1: Speaking Anxiety (Learner-Facing)}

\begin{quote}
\textbf{Task:} ``I can read, write and understand English well but I panic when I have to speak English, especially in front of other people. Why does this happen?''
\end{quote}

\begin{table}[h]
\centering
\small
\begin{tabular}{ll}
\toprule
\textbf{Attribute} & \textbf{Value} \\
\midrule
Role & Learner \\
Competency & C10 --- Manage Social-Emotional Aspects \\
Sub-competency & 10b \\
Resources & None \\
Context & --- \\
Context specificity & Minimal \\
\bottomrule
\end{tabular}
\end{table}

\textbf{Task Criteria:} Explains why speaking anxiety occurs (+8) $\cdot$ Provides strategies to manage speaking anxiety (+7)

\textbf{Consensus Criteria (10b):} 10b-01: Shows understanding and empathy (+7) $\cdot$ 10b-02: Raises awareness of self-efficacy (+5) $\cdot$ 10b-03: Develops self-regulated learning (+4)

\textbf{Universal Criteria:} 02u: Language appropriate for CEFR level (+9) $\cdot$ 06u: Response not appropriate for cultural sensitivities ($-10$)

\paragraph{Example 2: Lesson Planning with Resource (Teacher-Facing)}

\begin{quote}
\textbf{Task:} ``I've got this really cool text about the use of AI in music: [reading\_ai\_music\_b1.md]. I want to create a lesson for my B1 level teenagers. Can you help me plan a 45-minute lesson?''
\end{quote}

\begin{table}[h]
\centering
\small
\begin{tabular}{ll}
\toprule
\textbf{Attribute} & \textbf{Value} \\
\midrule
Role & Teacher \\
Competency & C02 --- Plan a Lesson \\
Sub-competency & 02a \\
Resources & reading\_ai\_music\_b1.md (B1-level reading, 250 words) \\
Context & CEFR B1; Upper-Secondary (age 15--18) \\
Context specificity & Partial \\
\bottomrule
\end{tabular}
\end{table}

\textbf{Task Criteria:} Creates a complete 45-minute lesson plan (+10) $\cdot$ Activities use the provided AI/music text (+8)

\textbf{Consensus Criteria (02a):} 02a-01: Includes appropriate pattern (PPP, ESA, TBLT) (+5) $\cdot$ 02a-02: Activities build knowledge/skills for goal (+6) $\cdot$ 02a-03: Clear structure for student profile (+8) $\cdot$ 02a-07: Realistic timings (+6) $\cdot$ 02a-09: Activities engage students (+8)

\textbf{Universal Criteria:} 02u: Language appropriate for CEFR level (+2, teacher-facing) $\cdot$ 03u: Response appropriate for learner characteristics (+5) $\cdot$ 05u: Response appropriate for learning context (+10, full context)

\subsection{Dataset Distribution}
\label{app:distribution}

The full benchmark is roughly evenly distributed across the 12 competencies. Tasks are annotated with context factors describing the pedagogical setting they instantiate; the distributions below characterise the coverage of these factors.

\begin{figure}[h]
\centering
\includegraphics[width=\columnwidth]{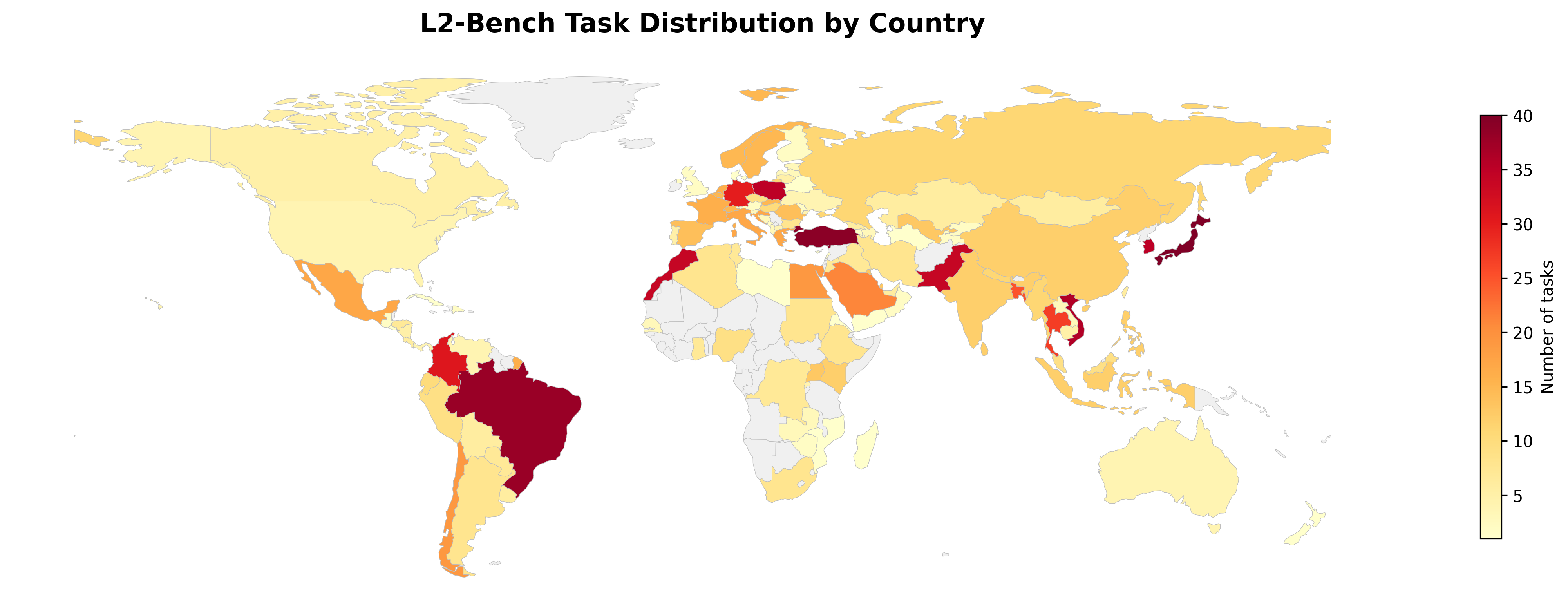}
\caption{Geographic distribution of L2-Bench tasks. Colour intensity encodes the number of country-specific tasks.}
\label{fig:task-country-map}
\end{figure}

\textbf{Geographic coverage.} Over 90\% of tasks carry a country-level context spanning 121 unique countries. The 70 highest-frequency countries where English is taught as a second language account for 86.9\% country-tagged tasks; the remaining are spread across 50 additional countries, giving long-tail coverage. Figure~\ref{fig:task-country-map} shows the choropleth distribution of task counts by country.

\textbf{Learner age.} An age group is specified for 88.8\% of tasks: Adult (32.7\%), Upper-Secondary (16.8\%), Lower-Secondary (15.1\%), Tertiary (13.2\%), Primary (10.6\%), and Pre-primary (0.4\%). Pre-primary is intentionally sparse and is omitted from the age-group performance breakdown (Table~\ref{tab:age_groups}).

\textbf{Stakeholder role.} Every task is tagged with the role of the person issuing the prompt: Teacher (71.1\%), Learner (9.1\%), Curriculum Designer (6.9\%), Guide/Trainer (5.5\%), Assessment Developer (2.9\%), and a small residue of mixed roles (Teacher \& Learner 3.4\%, Curriculum \& Teacher 0.9\%, Assessment \& Teacher 0.2\%). The performance breakdown in Appendix~I (Table~\ref{tab:role_performance}) reports the five single-role categories and omits the mixed combinations.

\textbf{Resource setting.} A combined resource level (aggregating economic context, materials, internet, and device availability) is specified for 25.6\% of tasks. Within the full dataset these split into High (15.1\%), Mixed (6.6\%), and Low (3.9\%) resource contexts; the performance breakdown (Table~\ref{tab:resource_levels}) reports these three levels.

\textbf{Context specification.} Every task is annotated for how fully its pedagogical context is described: Full (55.7\% of the dataset), Partial (33.1\%), and Minimal (11.2\%). This lets us quantify performance against the varying detail with which real-world prompts are specified.

\clearpage

\section{Practitioner Validation Study}
\label{app:validation}

This appendix provides full methodological detail for the practitioner validation study summarized in "L2-Bench construct validation" above.

\subsection{Overview and Research Objectives}

The Practitioner Validation Study assessed the validity of the L2-Bench dataset and scoring pipeline in an online study running from 2 to 12 March 2026 using education practitioners from across the world spanning 6 stakeholder groups (content developers, assessment specialists, teachers, generalist education professionals, academics, and learners) representing the dynamics of global pedagogy.

Our prior work \citet{edgell2026accuracyrobustevaluationmethodology} informed the approach to this study, where we set out to address the following four research objectives:
\begin{itemize}
    \item \textbf{RO1} Dataset validity: Practitioner agreement on task authenticity and criteria adequacy.
    \item \textbf{RO2} Answer quality: Practitioner preference for our reference vs.\ model-generated responses in blind A/B comparison.
    \item \textbf{RO3} Auto-scorer validity: Inter-judge agreement (IJA) between the LLM-as-a-Judge scores and practitioner scores against task rubrics.
    \item \textbf{RO4} Group differences: Whether ratings differ systematically across practitioner groups, competencies, or experience levels.
\end{itemize}

\subsection{Participant Recruitment and Allocation}
\label{app:recruitment}

Practitioners responded to a call for participants pre-study survey that was shared across 3 core sources: an institutional teacher panel ($n = 214$), external practitioner networks ($n = 102$), and internal staff ($n = 51$), yielding $N = 367$ allocated practitioners. All channels were closed networks for study integrity. Participants originating from the teacher panel were incentivized as per standard panel member compensation rates for one hour equivalent, while all other participation was voluntary (consistent with institutional compliance guidance). With the exception of internal staff (who were asked to contribute $\geq$3.5 hours total to the study), participants were asked to contribute 1 hour (exluding onboarding) equivalent to the study, however we incentivised those willing to contribute $\geq$3.5 hours total equivalent to the study by offering acknowledgement in a future publication(s). See Appendix~G.5 for further details on study ethics.

\begin{figure}[H]
\centering
\includegraphics[width=0.85\linewidth]{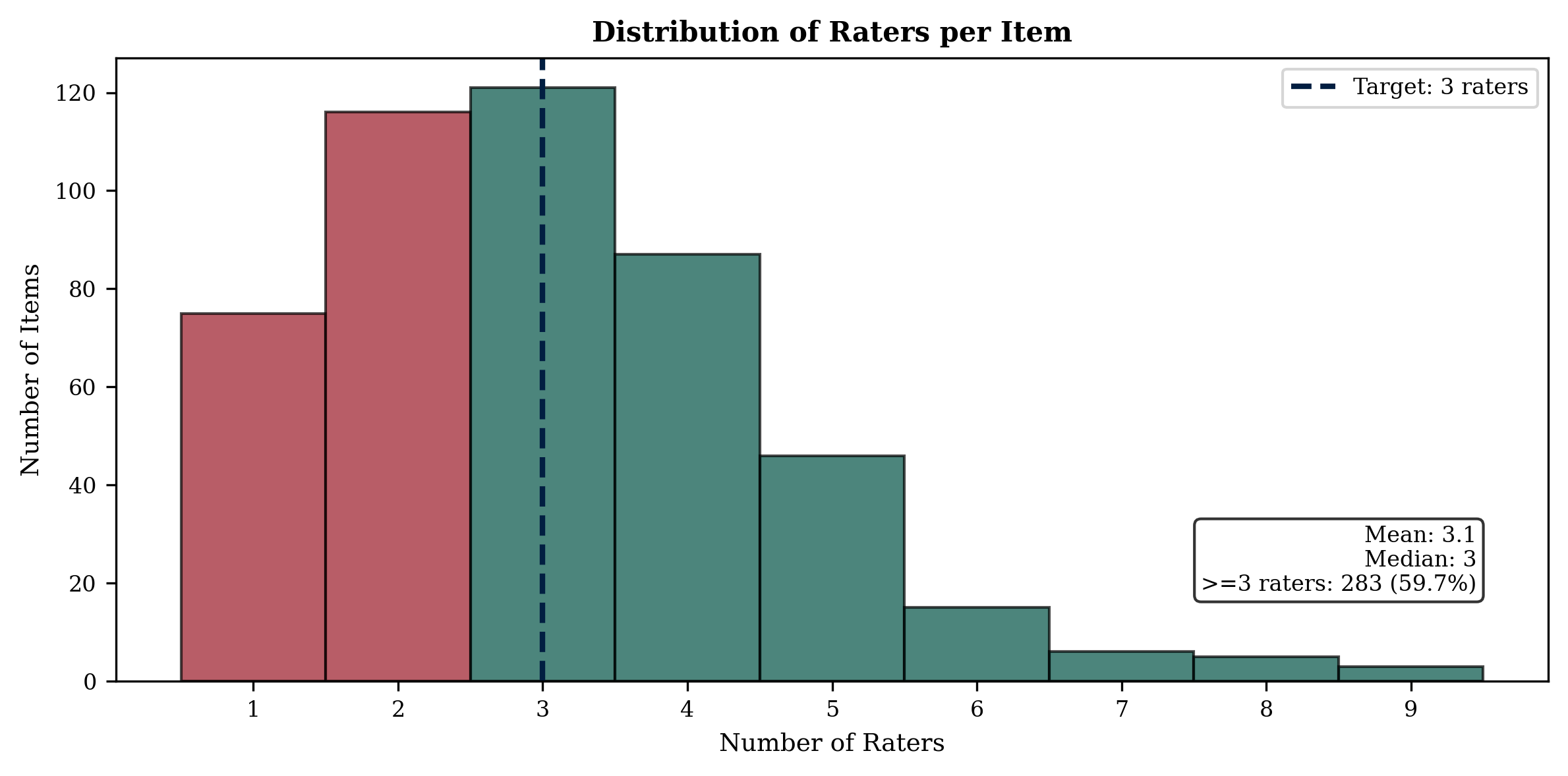}
\caption{Distribution of the number of independent practitioner ratings per item across the 474 rated items. The dashed line marks the target of three raters per item; bars meeting the target are shown in green.}
\label{fig:item-coverage}
\end{figure}

Communications were transparent and consistent across sources leading up to and throughout the study window. Approximately two thirds (241/367) of invited practitioners completed the study, with the dropout observed typical for voluntary online research. Despite having 241 practitioners participate during the study window, to ensure data quality, we excluded 20 raters who exhibited signals of systematic straight-lining or cheating using a two-tier quality exclusion protocol: (1) a hard rule for those  with impossibly fast responses (median time per item $<$1 min) and/or screenshot rates $\geq$50\%, and (2) a weighted quality score that accounted for: median time per item, fastest item, total time and screenshot rates; buffered by a leniency score that accounted for professional attributes: recruitment source, professional affiliation, experience, current role, ELT subject expertise and multiple expertise areas. This resulted in the $N = 221$ practitioners reported in the final study figures, who yielded 1,447 ratings across 474 items (see Table~\ref{tab:rater-summary} for a summary of the key rater statistics from the retained practitioner cohort, and Figure~\ref{fig:item-coverage} for the resulting per-item rater coverage).


\begin{table}[h]
\centering
\caption{Practitioner Validation Study rater summary statistics.}
\label{tab:rater-summary}
\small
\begin{tabular}{lc}
\toprule
\textbf{Metric} & \textbf{Value} \\
\midrule
Practitioners (post-exclusion) & 221 \\
Total ratings & 1,447 \\
Unique items rated & 474 \\
Mean / Median ratings per practitioner & 3.1 / 3 \\
Countries represented & 45 \\
\bottomrule
\end{tabular}
\end{table}

\paragraph{Practitioner Cohort} The L2-Bench practitioner validation study involved $N = 221$ practitioners across 45 countries, sourced from an internal teacher panel (54\%), external networks (31\%), and internal staff (15\%), where 85.5\% of participants self-reported $\geq$10 years of education experience, and 65\% currently worked as classroom teachers (63\%). Expertise included professional development trainers (24\%), assessment specialists (19\%), education researchers (14\%), and ensuring breadth across the profession (see Table~\ref{tab:demographics-source-exp} for details).

\begin{table}[h]
\centering
\caption{Practitioner demographics by source and experience.}
\label{tab:demographics-source-exp}
\small
\begin{tabular}{lcc|lcc}
\toprule
\textbf{Source} & \textbf{N} & \textbf{\%} & \textbf{Experience} & \textbf{N} & \textbf{\%} \\
\midrule
Teacher panel & 113 & 51.1\% & $>$10 years & 189 & 85.5\% \\
External & 73 & 33.0\% & 6--10 years & 18 & 8.1\% \\
Internal staff & 35 & 15.8\% & 2--5 years & 8 & 3.6\% \\
\bottomrule
\end{tabular}
\end{table}

\begin{figure}[H]
\centering
\includegraphics[width=\linewidth]{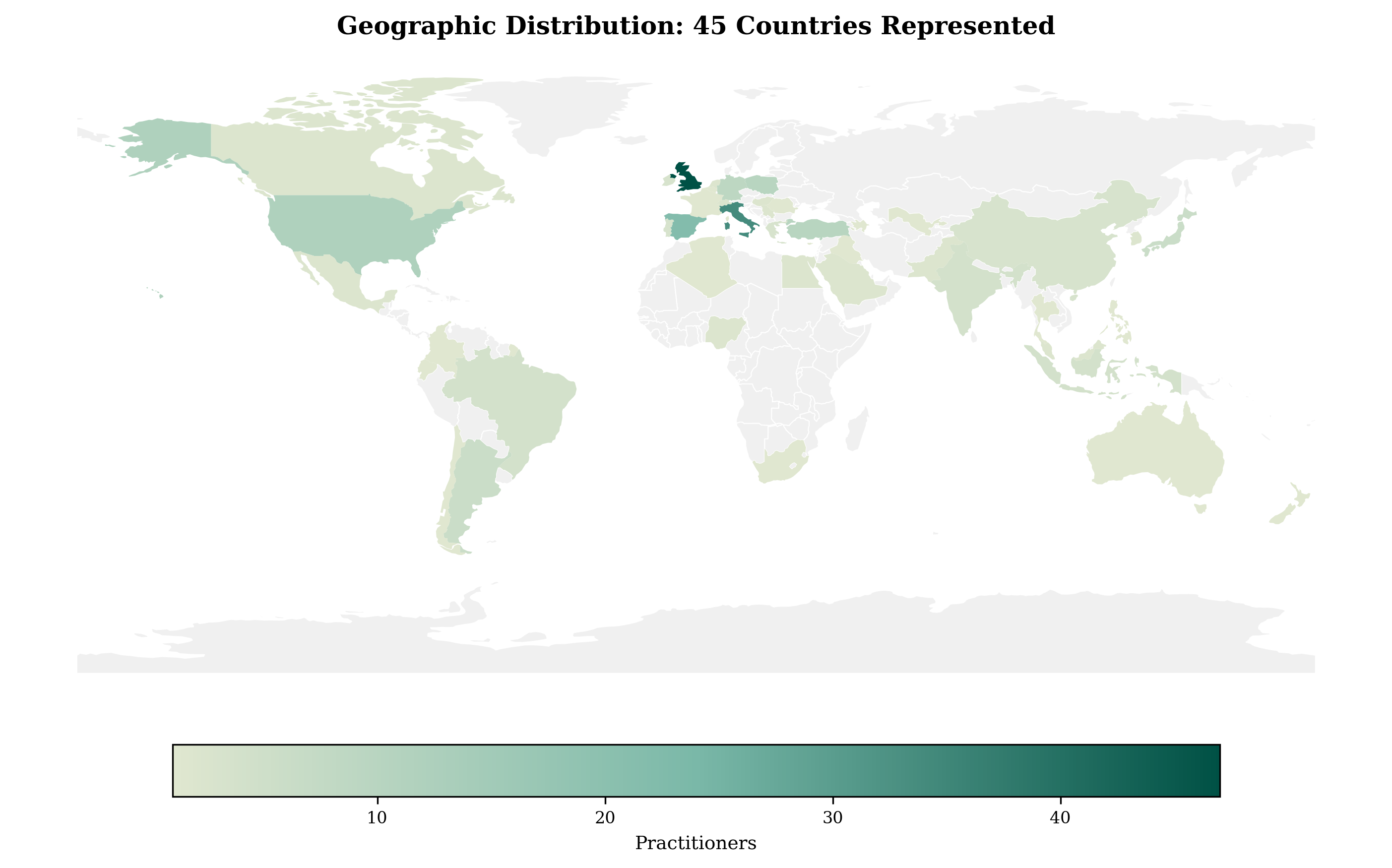}
\caption{Geographic distribution of the $N = 221$ retained practitioners across 45 countries. Shading intensity is proportional to the number of practitioners based in each country. While the cohort is anchored in Europe (Table~\ref{tab:geo-distribution}), the remaining practitioners span every populated continent.}
\label{fig:rater-world-map}
\end{figure}

The top 5 countries where the practitioner cohort are based are European, and account for just over half (55.3\%) of the practitioners in the study, as shown in Table~\ref{tab:geo-distribution}. Figure~\ref{fig:rater-world-map} shows the full geographic spread of the cohort, illustrating that the remaining practitioners are distributed across a further 40 countries spanning every populated continent.

\begin{table}[h]
\centering
\caption{Geographic distribution (top 10 countries) of practitioners.}
\label{tab:geo-distribution}
\small
\begin{tabular}{lcc|lcc}
\toprule
\textbf{Country} & \textbf{N} & \textbf{\%} & \textbf{Country} & \textbf{N} & \textbf{\%} \\
\midrule
United Kingdom & 47 & 21.3\% & United States & 9 & 4.1\% \\
Italy & 34 & 15.4\% & Japan & 6 & 2.7\% \\
Spain & 22 & 10.0\% & Argentina & 6 & 2.7\% \\
Poland & 10 & 4.5\% & Indonesia & 4 & 1.8\% \\
Germany & 9 & 4.1\% & Brazil & 4 & 1.8\% \\
\midrule
\multicolumn{3}{l}{Other (35 countries)} & \multicolumn{3}{c}{70 (31.7\%)} \\
\bottomrule
\end{tabular}
\end{table}

\subsection{Dataset Preparation and Item Allocation}

A stratified random sample of \textbf{504 items} (42 per competency) was drawn from the L2-Bench public dataset to target an average of three independent ratings per item, informed by extensive prior experimental design \citet{edgell2026accuracyrobustevaluationmethodology} and accounting for participant registration volumes. This sample represents approximately 50\% of the full public dataset; population inference is justified since all items are generated by the same hybrid human-AI authoring pipeline under consistent quality-assurance conditions (see Appendix~F).

Each item was formatted with the following schema: task input, task resources (inlined), model response (Claude Sonnet 4.6 via Amazon Bedrock), reference answer, evaluation criteria (task + consensus + universal consolidated), competency tag, sub-competency tag(s), and system prompt. Model responses were extracted from inspect-ai evaluation logs (see Appendix~B for open dataset logs).

A pre-study intake survey mapped each practitioner to competencies via a rule-based algorithm operating on 24 binary flags (current role, previous role, subject specialism, declared expertise, institutional research engagement). Two allocation caps applied: (1) assessment specialists ($n = 56$) were capped at $\leq$3 competencies so as to maximise their expertise on assessment competencies which were anticipated to have the least coverage; (2) all others capped at $\leq$8, with remaining slots filled via rarest-first priority. A minimum of two competencies per practitioner was enforced to keep participants engaged. Items were then randomly allocated from the study sample dataset according to these rules and practitioner time availability (see below).

\subsection{Study Platform and Procedure}

The study was delivered via a custom web application (AWS), enforcing a sequential time-gated workflow (Stages A, B, C) with stage progression locked until submission. Session events were recorded as an append-only log (51,975 events total), with study integrity supported by screenshot detection, event timings, and login-based access.

For study integrity, calibration materials (slides and a ~20min recording) were provided prior to and during the study window; an in-app checkbox was required to be checked on every login to the study platform to confirm that materials had been viewed.

For each task item shown to the practitioner, the validation workflow proceeded sequentially through 2 to 3 stages depending on their reported time availability:
\begin{itemize}
    \item \textbf{Stage A---Dataset Ratings ($\sim$10 min):} Rate task authenticity and criteria adequacy on 5-point Likert scales, with optional comments.
    \item \textbf{Stage B---Answer Preference ($\sim$5 min):} Choose a preferred response in a blind A/B comparison of two responses to the task presented in randomized order (one generated with Claude Sonnet 4.6, the other one was our reference answer).
    \item \textbf{Stage C---Rubric Scoring ($\sim$10 min):} After revealing the "AI response" (which is randomized to be either the AI response or our reference answer), practitioners score the "AI response" against the task rubric on each criterion as Pass/Fail.
\end{itemize}

\begin{figure}[p]
\centering
\includegraphics[width=0.62\linewidth]{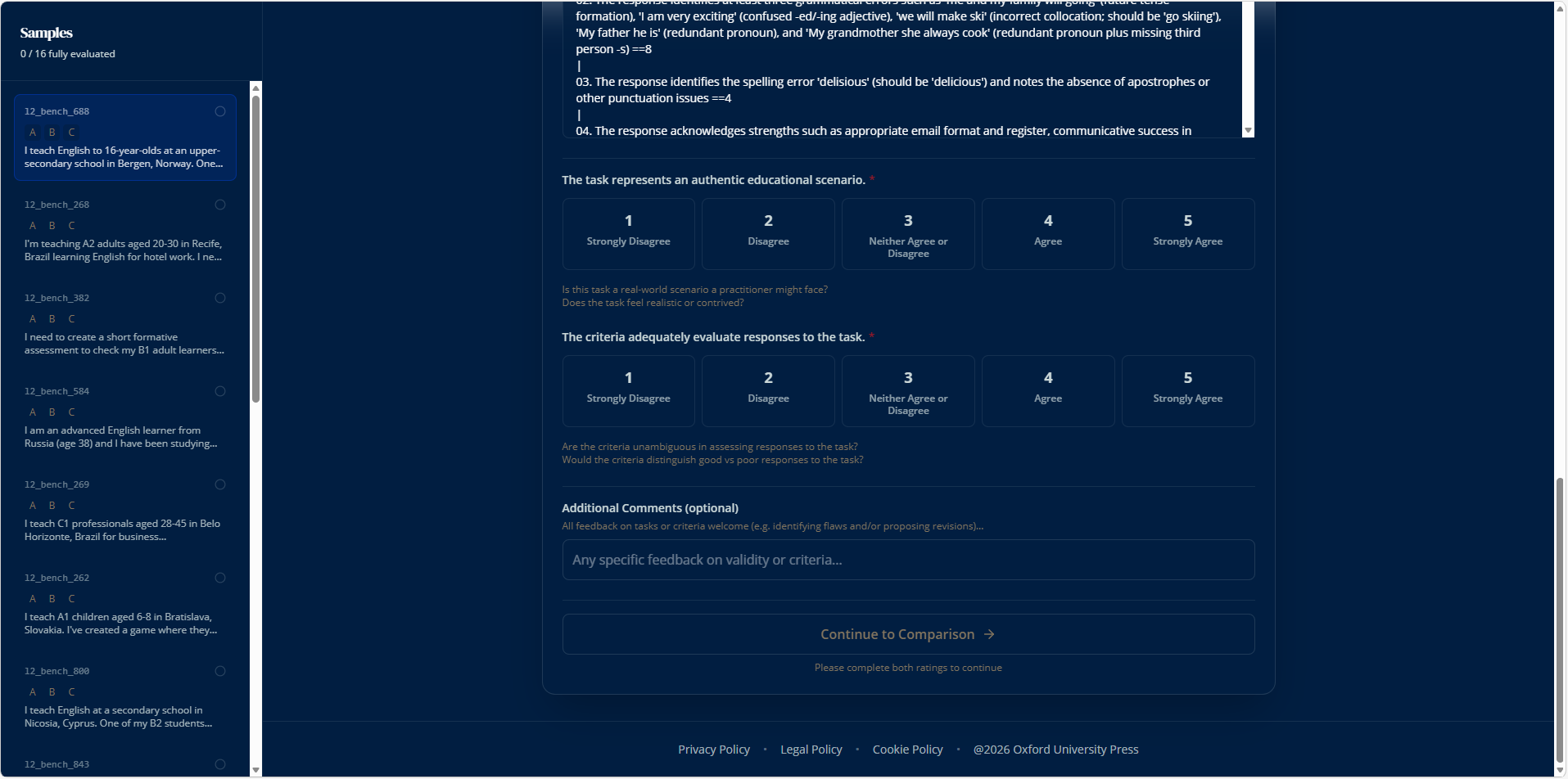}\\[0.3em]
\includegraphics[width=0.62\linewidth]{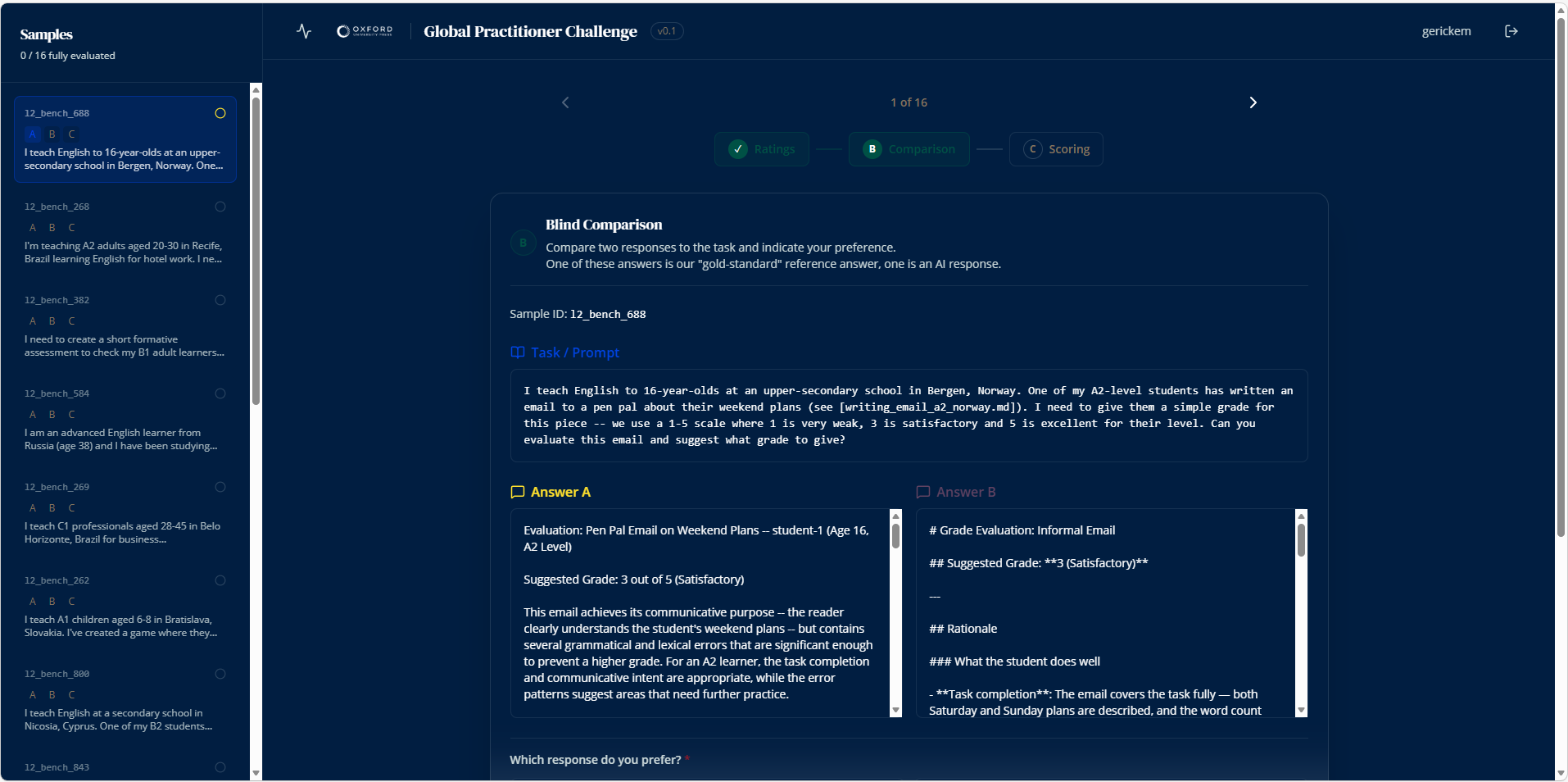}\\[0.3em]
\includegraphics[width=0.62\linewidth]{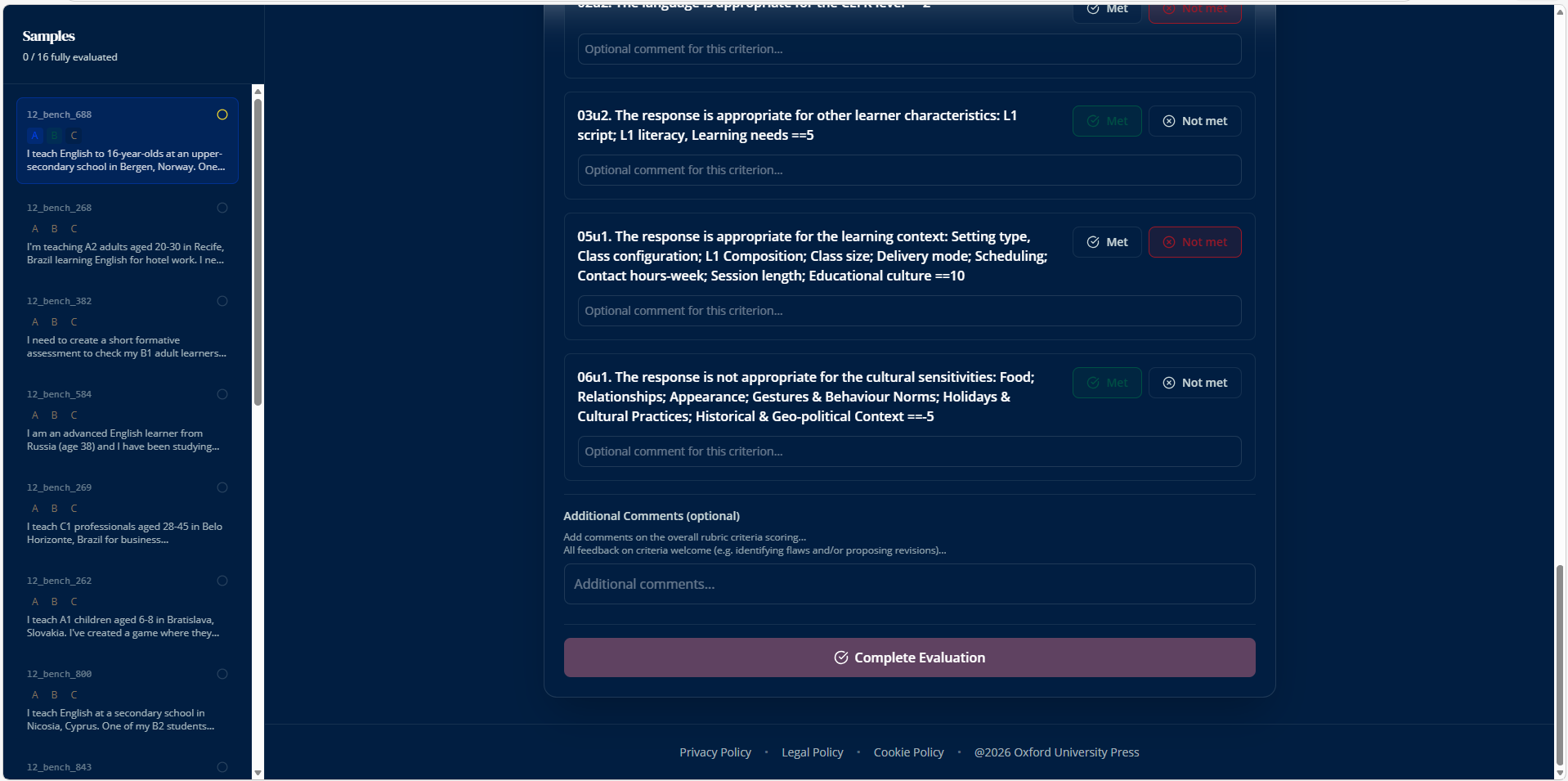}\\[0.3em]
\includegraphics[width=0.62\linewidth]{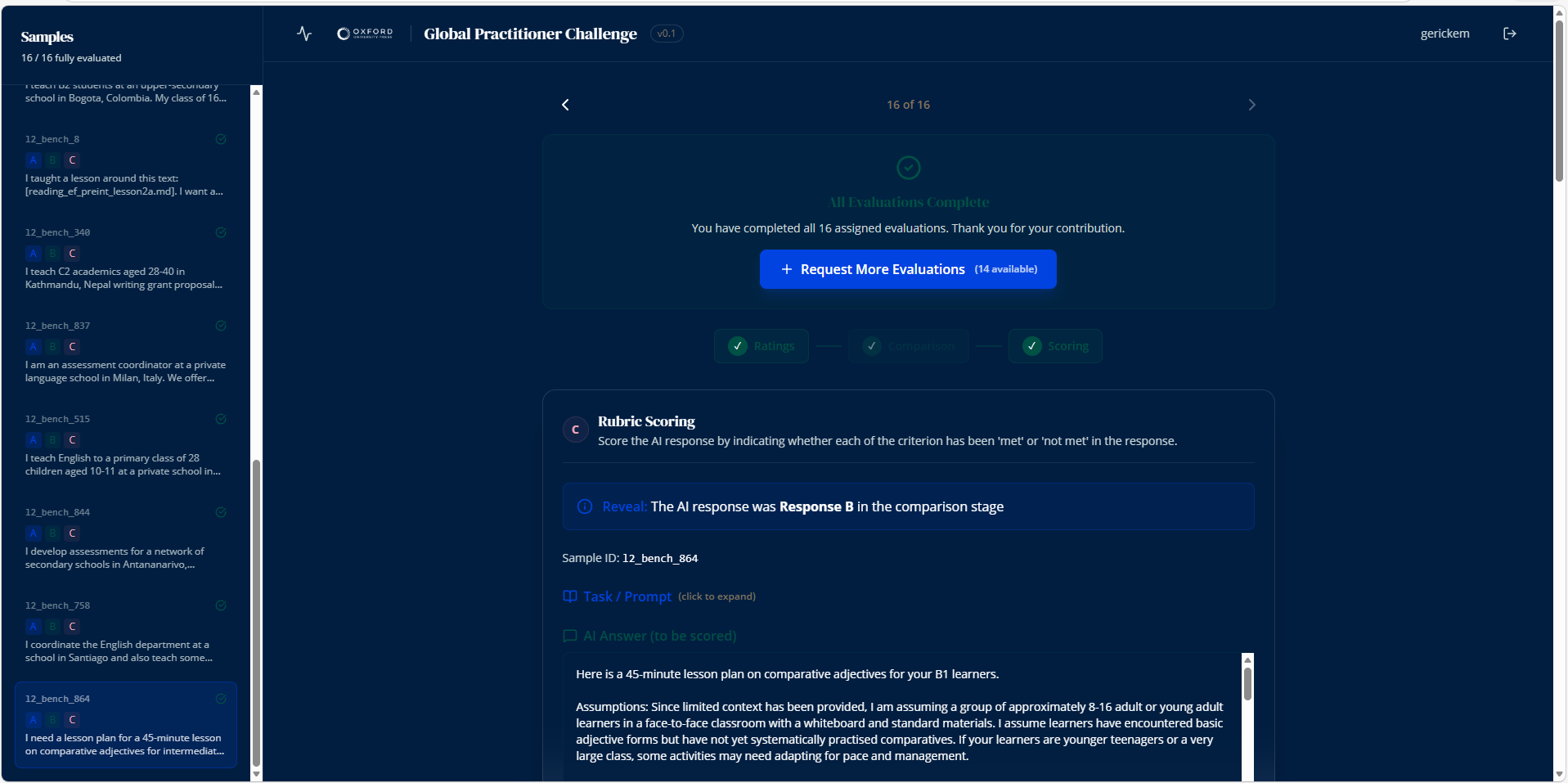}
\caption{The practitioner study platform, shown in study-procedure order. Top: Stage A dataset rating screen, where practitioners rate task authenticity and criteria adequacy on 5-point Likert scales. Second: Stage B blind A/B answer-preference comparison screen. Third: Stage C per-criterion Pass/Fail rubric-scoring screen. Bottom: item completion screen confirming submission, with option to request a batch of additional items.}
\label{fig:app-screens}
\end{figure}

Practitioners committing 1 hour completed Stages A+B only (4 items, $\sim$15 min each); those committing $\geq$3.5 hours completed Stages A+B+C ($\sim$25 min each). Optionally, practitioners could request additional items (Stages A+B+C), released as blocks of 6 items, with a cap of up to 30 items total. Figure~\ref{fig:app-screens} shows four key screens of the study platform that illustrate the study procedure.

\subsection{Study Ethics}

The study was conducted under institutional oversight. All data infrastructure was reviewed and approved by privacy and cybersecurity teams. No personally identifiable information is present in the analytical datasets; usernames are pseudonymized. All data handling complies with UK GDPR. Research ethics were governed by Oxford University Press. A supplementary IRB was initiated under the auspices of the Oxford Internet Institute (University of Oxford) departmental research ethics committee, but was ruled exempt.

Practitioners affirmed informed consent via an in-app checkbox on first login to the study platform; study terms were available from the first communication when registering interest, and were additionally published on a linked website that remained available throughout the study, with transparent communications sent throughout. The blind A/B comparison design (Stage B of the study) was a methodological necessity for measuring genuine preference; participants were informed of this design in post-study communications and given option to withdraw.

\subsection{Statistical Methods}
\label{app:stats}

To answer our research objectives, statistical methods built upon prior work \citet{edgell2026accuracyrobustevaluationmethodology}, noting that the study data collection resulted in sparse coverage (only 60\% of items achieved the $\geq$3 rater target) that limited the practicability of some previously registered statistical methods. We therefore recommend that future validations implement dynamic item allocations, for example releasing only 1 batch of items at a time until they reach the pre-determined rater target before releasing the next batch.

Statistical analyses are organized by research objective:
\begin{itemize}
    \item \textbf{RO1 (Dataset validity):} Mean scores per competency; one-sample $t$-tests against targets (4.0/5 authenticity, 3.5/5 criteria adequacy); IAA via Krippendorff's alpha; IIC via Cronbach's alpha; rater severity via ICC from mixed-effects models; authenticity-criteria gap via Wilcoxon signed-rank.
    \item \textbf{RO2 (Answer quality):} One-sample binomial test (target 70\% reference preference, $p < 0.05$).
    \item \textbf{RO3 (Auto-scorer validity):} Cohen's kappa (target $\geq 0.60$); recall target $\geq 0.80$.
    \item \textbf{RO4 (Group differences):} Mann-Whitney $U$ pairwise comparisons; power approximately 68--92\% depending on group sizes.
\end{itemize}

\textbf{Krippendorff's Alpha (IAA)}. Used instead of Fleiss' kappa because sparse coverage ($\sim$3 raters/item) produces incomplete matrices:
\begin{equation}
\alpha = 1 - \frac{D_o}{D_e}
\label{eq:kripp-alpha}
\end{equation}
where $D_o$ is observed disagreement and $D_e$ is expected disagreement under chance, with ordinal distance weights $\delta_{ck} = |c - k|$. Thresholds: $\alpha \geq 0.80$ reliable; 0.667--0.80 tentative; $< 0.667$ unreliable.

\textbf{Cronbach's Alpha (IIC)}:
\begin{equation}
\alpha = \frac{k}{k-1} \left(1 - \frac{\sum_{i=1}^{k} \sigma^2_i}{\sigma^2_T}\right)
\label{eq:cronbach-alpha}
\end{equation}
Thresholds: $\geq 0.90$ excellent; 0.80--0.90 good; 0.70--0.80 acceptable; 0.60--0.70 questionable; 0.50--0.60 poor.

\textbf{Intraclass Correlation (ICC)}:
\begin{equation}
\text{ICC} = \frac{\sigma^2_{\text{rater}}}{\sigma^2_{\text{rater}} + \sigma^2_{\text{resid}}}
\label{eq:icc}
\end{equation}
Interpretation: $<$ 0.10 negligible, 0.10--0.30 small, 0.30--0.50 moderate, $\geq$ 0.50 large.

\textbf{Cohen's Kappa (IJA)} (LLM-Judge vs.\ human majority):
\begin{equation}
\kappa = \frac{P_o - P_e}{1 - P_e}
\label{eq:cohen-kappa}
\end{equation}
Thresholds: $< 0$ poor; 0--0.20 slight; 0.21--0.40 fair; 0.41--0.60 moderate; 0.61--0.80 substantial; 0.81--1.00 almost perfect.

\textbf{Auto-scorer sensitivity (recall)}. We prioritise recall for detecting failures, since a false negative (passing a poor response) risks exposing learners to inadequate content, whereas a false positive is caught by human review:
\begin{equation}
\text{Sensitivity} = \frac{TP}{TP + FN}
\label{eq:recall}
\end{equation}

\textbf{Binomial test (A/B preference)}. For blind reference-vs-model response comparisons we use a one-tailed binomial test against the null of no preference:
\begin{equation}
p\text{-value} = \sum_{x=k}^{n} \binom{n}{x} p_0^x (1-p_0)^{n-x}
\label{eq:binomial}
\end{equation}
where $k$ is observed reference preferences, $n$ is total trials and $p_0 = 0.50$.

\textbf{Variance-decomposition mixed-effects model}. To separate stable rater-severity effects from item-level variation in the ratings, we fit a linear mixed-effects model with a random rater intercept:
\begin{equation}
y_{ij} = \beta_0 + u_i + \varepsilon_{ij}
\label{eq:mixed-effects}
\end{equation}
where $y_{ij}$ is the rating by rater $i$ on item $j$, $\beta_0$ the grand intercept, $u_i \sim N(0, \sigma_{\text{rater}}^2)$ the rater random intercept (capturing systematic severity offsets), and $\varepsilon_{ij} \sim N(0, \sigma_{\text{resid}}^2)$ the residual (item quality plus error). The intraclass correlation ICC $= \sigma_{\text{rater}}^2 / (\sigma_{\text{rater}}^2 + \sigma_{\text{resid}}^2)$ then quantifies the share of variance attributable to rater severity; results are reported in Appendix~G, Construct Validation Results.

\textbf{Group comparisons (rank-biserial)}. For group contrasts, we use Mann-Whitney $U$ tests on per-rater means, reporting the rank-biserial correlation as the effect size:
\begin{equation}
r = \frac{2U}{n_1 n_2} - 1
\label{eq:rank-biserial}
\end{equation}
where $U$ is the Mann-Whitney statistic and $n_1, n_2$ the group sizes.

\textbf{Bootstrap confidence intervals}. For statistics without closed-form sampling distributions (Krippendorff's $\alpha$, Cronbach's $\alpha$), we use percentile bootstrap resampling ($B = 500$ iterations); the $100(1-\gamma)\%$ interval is
\begin{equation}
\text{CI} = \left[\, \hat{\theta}^{*}_{(\gamma/2)},\; \hat{\theta}^{*}_{(1-\gamma/2)} \,\right]
\label{eq:bootstrap-ci}
\end{equation}
where $\hat{\theta}^{*}_{(q)}$ is the $q$-th quantile of the bootstrap replicate statistics. (The standard-error framework used for the aggregate \emph{model leaderboard} scores is a distinct benchmark-scoring statistic and is described separately in Appendix~I.2.)

\subsection{Construct Validation Results}
\label{app:construct-validation-results}

This subsection reports the evidence bearing on construct validity.

\begin{figure}[t]
\centering
\includegraphics[width=\linewidth]{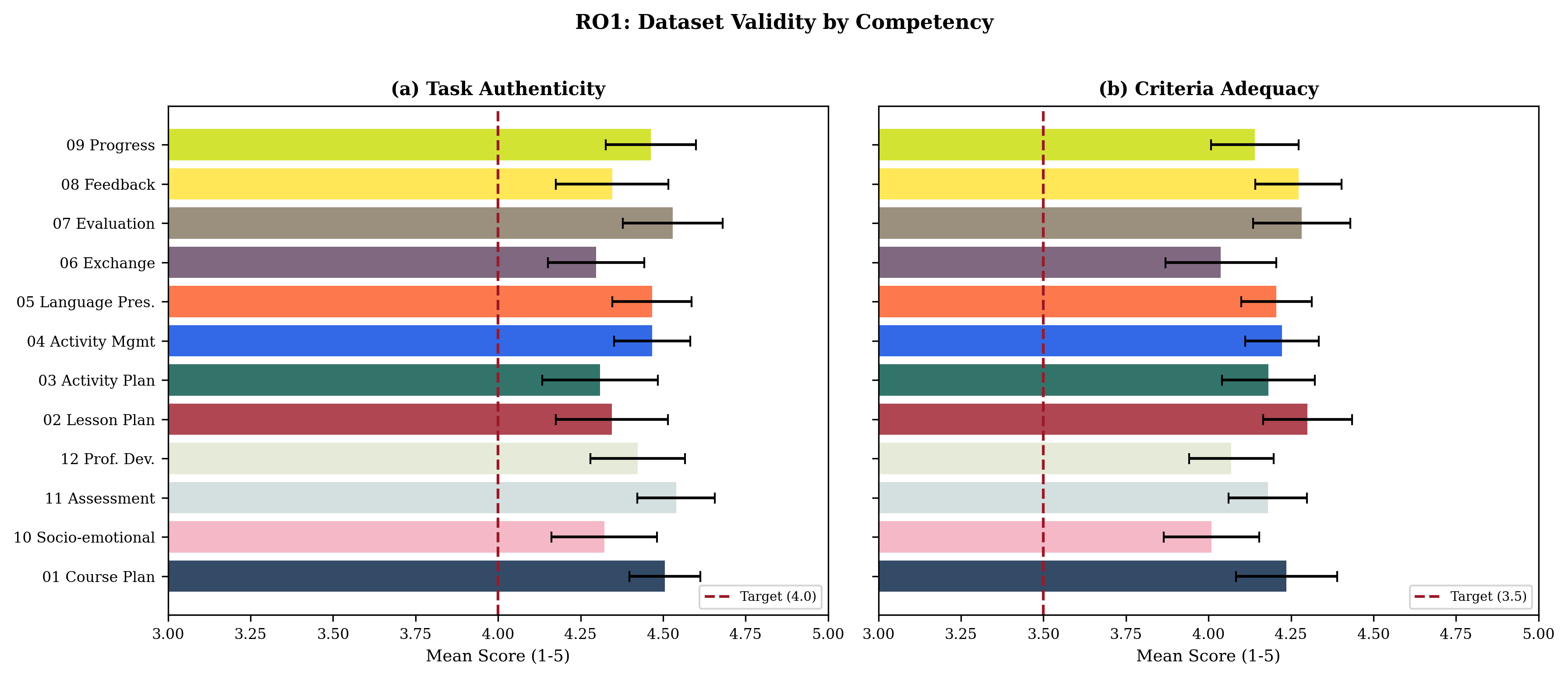}
\caption{Practitioner ratings of (a) task authenticity and (b) criteria adequacy, by competency.}
\label{fig:validity-auth-crit}
\end{figure}

Both task authenticity ($M = 4.42$, 95\% CI [4.38, 4.46]) and criteria adequacy ($M = 4.18$, 95\% CI [4.14, 4.22]) significantly exceeded their targets of 4.0 and 3.5 respectively at an overall level, and across all 12 competencies individually (see Figure~\ref{fig:validity-auth-crit}), providing strong evidence that the construct is coherent at the sub-competency level.

\paragraph{Variance decomposition.} To disentangle rater severity from item-level variation, we fit the variance-decomposition mixed-effects model of Equation~\ref{eq:mixed-effects} (a random-rater-intercept model). Table~\ref{tab:variance-decomposition} shows that 27--33\% of variance is attributable to rater severity---consistent with documented patterns in educational assessment. The remaining 67--73\% is residual (item quality + error), providing evidence that practitioners track a shared underlying quality signal while differing in their absolute use of the rating scale.

\begin{table}[h]
\centering
\caption{Variance decomposition results.}
\label{tab:variance-decomposition}
\small
\begin{tabular}{lcccc}
\toprule
\textbf{Measure} & \textbf{Rater Var.} & \textbf{Residual Var.} & \textbf{ICC} & \textbf{\% Rater} \\
\midrule
Task Authenticity & 0.158 & 0.436 & 0.265 & 27\% \\
Criteria Adequacy & 0.163 & 0.340 & 0.325 & 33\% \\
\bottomrule
\end{tabular}
\end{table}



\paragraph{Group differences.} Practitioners from diverse professional backgrounds rated the benchmark items similarly. Mann-Whitney $U$ tests comparing ratings across source channels, experience levels, and geographic regions found no effect sizes exceeding $r = 0.10$, supporting construct validity across the target user population. \emph{By source}, external practitioners ($n = 73$) showed marginally higher authenticity ($M = 4.52$) than internal staff ($n = 35$; $M = 4.28$; $U = 108{,}961$, $p = 0.003$, $r = 0.08$) and the teacher panel ($n = 113$; $M = 4.42$; $p = 0.001$, $r = 0.09$), though all effect sizes were small ($r < 0.10$). \emph{By experience}, no significant differences emerged between $>$10 years ($n = 189$), 6--10 years ($n = 18$), or 2--5 years ($n = 8$) on either measure (all $p > 0.05$, negligible effects). \emph{By geography}, Japan ($n = 30$ ratings) showed lower task authenticity ($M = 3.71$; $p < 0.001$, $r = 0.09$); individual country samples were too small for robust inference, and the patterns are consistent with sampling variation rather than systematic cultural bias.

\subsection{Answer Preference Results}
\label{app:preference-results}

In a blind A/B preference comparison of responses, practitioners preferred our reference answer 51.3\% of the time versus Claude Sonnet 4.6 responses, although this was not significant. More critically, the result falls far short of the 70\% preference target, with only 1 of 12 competencies showing significant preference (see Table~\ref{tab:preference-comparison}).

Table~\ref{tab:preference-comparison} presents the blind preference comparison results by competency.

\begin{table}[h]
\centering
\caption{Blind preference comparison by competency.}
\label{tab:preference-comparison}
\small
\begin{tabular}{lrrrrl}
\toprule
\textbf{Competency} & \textbf{N} & \textbf{Ref \%} & \textbf{AI \%} & \textbf{p-value} & \textbf{Sig} \\
\midrule
01 -- Course Planning & 108 & 52.8 & 47.2 & 0.588 & \\
02 -- Lesson Planning & 104 & 51.0 & 49.0 & 0.922 & \\
03 -- Activity Planning & 112 & 44.6 & 55.4 & 0.167 & \\
04 -- Activity Management & 126 & 50.8 & 49.2 & 0.879 & \\
05 -- Language Presentation & 118 & 50.0 & 50.0 & 1.000 & \\
06 -- Exchange Partner & 114 & 53.5 & 46.5 & 0.426 & \\
07 -- Performance Evaluation & 130 & 48.5 & 51.5 & 0.638 & \\
08 -- Giving Feedback & 124 & 54.0 & 46.0 & 0.361 & \\
09 -- Progress Tracking & 118 & 56.8 & 43.2 & 0.152 & \\
10 -- Emotional Intelligence & 100 & 64.0 & 36.0 & 0.004 & * \\
11 -- Assessment Creation & 106 & 47.2 & 52.8 & 0.588 & \\
12 -- Professional Development & 142 & 49.3 & 50.7 & 0.872 & \\
\midrule
\textbf{Overall} & \textbf{1,402} & \textbf{51.3} & \textbf{48.7} & \textbf{0.188} & \\
\bottomrule
\end{tabular}
\begin{flushleft}
\scriptsize Note: * = $p < 0.05$ (two-tailed exact binomial test against 50\% null).
\end{flushleft}
\end{table}

Only Competency 10 (socio-emotional aspects) showed significant preference for reference answers (64.0\% vs.\ 36.0\%), providing evidence that practitioners detect quality differences specifically in tasks requiring nuanced interpersonal sensitivity. The non-significant results for the remaining eleven competencies are noteworthy as they suggest that practitioners cannot distinguish frontier model responses from expert-reviewed pedagogical exemplars through blind evaluation, however we caveat that our experimental design did not add a "no preference" response, which would have allowed for a more concrete interpretation.

\clearpage

\section{Judge Building}
\label{app:judge}

This appendix provides the full methodological detail for the automated LLM-as-a-Judge scoring pipeline optimization summarized in Section 3.

\subsection{Judge Prompt Design}

Four prompt variants were developed to ablate two orthogonal design dimensions---reference guidance and prompt scaffolding (chain-of-thought and few-shot examples, which we refer to as "CoT")---producing a 2 $\times$ 2 factorial shown in Table~\ref{tab:judge-prompts}:

\begin{table}[h]
\centering
\caption{Judge prompt variants.}
\label{tab:judge-prompts}
\small
\begin{tabular}{llp{6cm}}
\toprule
\textbf{Version} & \textbf{Guidance / Mode} & \textbf{Description} \\
\midrule
v1 & Ref-guided classifier & Reference answer supplied; judge returns Pass/Fail directly \\
v2 & Ref-guided CoT & Reference answer supplied; judge produces chain-of-thought before verdict \\
v3 & Plain classifier & No reference answer; judge returns Pass/Fail directly \\
v4 & Plain CoT & No reference answer; judge produces chain-of-thought before verdict \\
\bottomrule
\end{tabular}
\end{table}

The reference-guided variants (v1, v2) represent the standard pipeline described in \citet{edgell2026accuracyrobustevaluationmethodology}: the judge receives the expert reference answer alongside the task input, rubric criterion, and model response. The plain variants (v3, v4) test whether the pipeline remains effective without reference guidance.

\subsubsection{Judge prompt text.}
\label{sec:judge-prompt-text}
All prompts grade one rubric criterion at a time, returning a single Pass/Fail verdict; the fields \texttt{\{task\_text\}}, \texttt{\{reference\_answer\}} (v1, v2 only), \texttt{\{ai\_response\}}, and \texttt{\{criterion\_description\}} are interpolated at scoring time. On a malformed verdict the pipeline retries once with the suffix ``\emph{Error: You output an invalid value. You MUST output 'true' or 'false' ONLY in your response.}'' The production judge uses the v1 reference-guided classifier, reproduced below.

\begin{lstlisting}[basicstyle=\ttfamily\scriptsize,breaklines=true,frame=single,caption={v1 reference-guided classifier prompt (production judge).}]
You are an expert evaluator for AI-generated responses in second language (L2) education.
Your job is to judge whether an AI response meets a single scoring criterion.

## Original Task
{task_text}

## Reference Answer
{reference_answer}

## AI Response
{ai_response}

## Criterion
- Description: {criterion_description}

## Instructions
Determine whether the AI response meets the criterion above.
Use the reference answer as a guide for what a high-quality response looks like, but note that alternative valid approaches may exist.

- 'true' if the criterion IS met.
- 'false' if the criterion IS NOT met.

You MUST output either 'true' or 'false' and nothing else. Otherwise your answer will NOT be parsed correctly.
\end{lstlisting}

The chain-of-thought variants (v2, v4) share the same header but replace the terminal instruction with a request for a written critique before the verdict, and prepend explicit Pass/Fail definitions plus three in-context worked examples (one Pass, one Fail, one borderline Pass). The plain variants (v3, v4) omit the \texttt{\{reference\_answer\}} block. The v3 plain classifier scaffold is:

\begin{lstlisting}[basicstyle=\ttfamily\scriptsize,breaklines=true,frame=single,caption={v3 plain (no-reference) chain-of-thought scaffold.}]
You are an expert evaluator for AI-generated responses in second language (L2) education.
Your job is to judge whether an AI response meets a single scoring criterion.

## Original Task
{task_text}

## AI Response
{ai_response}

## Criterion
- Description: {criterion_description}

## Instructions
First, write a detailed critique that:
- Identifies whether the specific requirement of the criterion is present.
- Cites specific evidence from the AI response to support your judgement.

After your critique, output your final verdict on a new line.
Output ONLY 'true' if the criterion IS met, or ONLY 'false' if the criterion IS NOT met.
Do not output any other text after your verdict - it will NOT be parsed correctly.
\end{lstlisting}

\subsection{Judge Performance Experiment}
\label{app:judge-tuning}

The judge performance experiment drew on 48 tasks (4 per competency) from the Practitioner Validation Study sample that had matched human criterion-level scores, enabling direct comparison between automated and human verdicts. Judge experiments were conducted with AWS Bedrock and inspect-ai pipelines, with temperature set to 0 where possible to ensure deterministic, reproducible outputs (configurations can be found in Table~\ref{tab:judge-tuning-config}; see Appendix~B for open-source dataset):

\begin{table}[H]
\centering
\caption{Judge performance experiment configuration.}
\label{tab:judge-tuning-config}
\small
\begin{tabular}{lp{8cm}}
\toprule
\textbf{Setting} & \textbf{Value} \\
\midrule
Judge temperature & 0, otherwise cloud provider default \\
Judge reasoning\_tokens & 1,024 \\
Judge max\_retries & 0 \\
Inference method & AWS Bedrock Batch Inference (Converse API) \\
Tasks & 48 (4 per competency) \\
Criteria & 875 \\
\bottomrule
\end{tabular}
\end{table}

The judge performance experiment was split into two parts:

\textbf{Part A---Judge-Human Alignment.} Three judge models (Claude Sonnet 4.5, Claude Sonnet 4.6, DeepSeek V3.2) each evaluated the same Sonnet 4.6 solver responses that human raters scored in the Practitioner Validation Study. Three prompt versions (v1, v2, v3) were tested, yielding 9 configurations. Each configuration produced 875 criterion-level verdicts.

\textbf{Part B---Reference Answer Accuracy.} The same 48 tasks were re-scored, this time with the judge evaluating the expert reference (gold standard) answer. Three judge models $\times$ 2 prompt versions (v3, v4---plain only) yielded 6 configurations.

\textbf{Total:} 15 configurations across Parts A and B, generating approximately 13,125 criterion-level judge scoring records.

\begin{table}[H]
\centering
\caption{Part A epoch-to-configuration mapping (Judge vs Human on AI Solver Output).}
\label{tab:part-a-epochs}
\small
\begin{tabular}{clp{5cm}}
\toprule
\textbf{Epoch} & \textbf{Judge Model} & \textbf{Prompt Version} \\
\midrule
1 & Claude Sonnet 4.5 & v1 (ref-guided classifier) \\
2 & Claude Sonnet 4.6 & v1 (ref-guided classifier) \\
3 & DeepSeek V3.2 & v1 (ref-guided classifier) \\
4 & Claude Sonnet 4.5 & v2 (ref-guided CoT) \\
5 & Claude Sonnet 4.6 & v2 (ref-guided CoT) \\
6 & DeepSeek V3.2 & v2 (ref-guided CoT) \\
7 & Claude Sonnet 4.5 & v3 (plain classifier) \\
8 & Claude Sonnet 4.6 & v3 (plain classifier) \\
9 & DeepSeek V3.2 & v3 (plain classifier) \\
\bottomrule
\end{tabular}
\end{table}

\begin{table}[H]
\centering
\caption{Part B epoch-to-configuration mapping (Judge vs Expected on Reference Answers).}
\label{tab:part-b-epochs}
\small
\begin{tabular}{clp{5cm}}
\toprule
\textbf{Epoch} & \textbf{Judge Model} & \textbf{Prompt Version} \\
\midrule
1 & Claude Sonnet 4.5 & v3 (plain classifier) \\
2 & Claude Sonnet 4.6 & v3 (plain classifier) \\
3 & DeepSeek V3.2 & v3 (plain classifier) \\
4 & Claude Sonnet 4.5 & v4 (plain CoT) \\
5 & Claude Sonnet 4.6 & v4 (plain CoT) \\
6 & DeepSeek V3.2 & v4 (plain CoT) \\
\bottomrule
\end{tabular}
\end{table}

\subsubsection{Judge Performance Results.}
The full results from Part~A of the judge performance experiment can be seen in the following table:

\begin{table}[H]
\centering
\caption{Part A --- Judge-Human Alignment (sorted by F1).}
\label{tab:judge-human-alignment}
\small
\begin{tabular}{llccccc}
\toprule
\textbf{Judge Model} & \textbf{Prompt} & \textbf{Acc.} & \textbf{Prec.} & \textbf{Rec.} & \textbf{F1} & \textbf{$\kappa$} \\
\midrule
DeepSeek V3.2 & v1 (ref-guided) & 91.1\% & 0.943 & 0.941 & 0.942 & 0.746 \\
Sonnet 4.6 & v1 (ref-guided) & 90.2\% & 0.935 & 0.938 & 0.936 & 0.719 \\
DeepSeek V3.2 & v3 (plain) & 89.4\% & 0.928 & 0.934 & 0.931 & 0.698 \\
Sonnet 4.5 & v1 (ref-guided) & 88.7\% & 0.921 & 0.929 & 0.925 & 0.681 \\
Sonnet 4.6 & v3 (plain) & 88.1\% & 0.917 & 0.923 & 0.920 & 0.664 \\
Sonnet 4.5 & v3 (plain) & 86.9\% & 0.908 & 0.915 & 0.911 & 0.631 \\
DeepSeek V3.2 & v2 (CoT) & 85.4\% & 0.896 & 0.903 & 0.899 & 0.594 \\
Sonnet 4.6 & v2 (CoT) & 84.2\% & 0.887 & 0.891 & 0.889 & 0.561 \\
Sonnet 4.5 & v2 (CoT) & 82.8\% & 0.874 & 0.879 & 0.876 & 0.518 \\
\bottomrule
\end{tabular}
\end{table}

Key findings from Part A:
\begin{itemize}
    \item \textbf{Best human alignment:} DeepSeek V3.2 with v1 (reference-guided classifier) achieved F1 = 0.942, Cohen's $\kappa = 0.746$, accuracy = 91.1\%.
    \item \textbf{Model ranking:} DeepSeek V3.2 $>$ Sonnet 4.6 $>$ Sonnet 4.5 across all prompt versions on F1.
    \item \textbf{Prompt ranking:} v1 (reference-guided classifier) $\geq$ v3 (plain classifier) $>$ v2 (reference-guided CoT).
\end{itemize}

The full results from part B of the judge performance experiment can be seen in the following table:

\begin{table}[H]
\centering
\caption{Part B --- Reference Answer Accuracy (sorted by accuracy).}
\label{tab:ref-answer-accuracy}
\small
\begin{tabular}{llccccc}
\toprule
\textbf{Judge Model} & \textbf{Prompt} & \textbf{Acc.} & \textbf{Prec.} & \textbf{Rec.} & \textbf{F1} & \textbf{$\kappa$} \\
\midrule
Sonnet 4.6 & v3 (plain) & 89.9\% & 0.941 & 0.935 & 0.938 & 0.672 \\
Sonnet 4.5 & v3 (plain) & 88.2\% & 0.928 & 0.921 & 0.924 & 0.629 \\
DeepSeek V3.2 & v3 (plain) & 87.1\% & 0.919 & 0.913 & 0.916 & 0.598 \\
Sonnet 4.6 & v4 (CoT) & 85.8\% & 0.907 & 0.898 & 0.902 & 0.561 \\
Sonnet 4.5 & v4 (CoT) & 84.3\% & 0.893 & 0.884 & 0.888 & 0.519 \\
DeepSeek V3.2 & v4 (CoT) & 82.7\% & 0.879 & 0.867 & 0.873 & 0.472 \\
\bottomrule
\end{tabular}
\end{table}

\subsection{Judge Stability Experiment.}
\label{app:stability}

To assess scoring determinism, we conducted a 3-resample stability test: five tasks were scored three times independently by each of 16 model $\times$ prompt configurations (4 models $\times$ 4 prompt versions (see Table~\ref{tab:stability-config} below for configurations), including Kimi K2.5), with total of 1,216 criterion-level stability observations were collected, allowing us to measure how often the binary criterion may "flip".

\begin{table}[H]
\centering
\caption{Stability experiment configuration.}
\label{tab:stability-config}
\small
\begin{tabular}{lp{8cm}}
\toprule
\textbf{Setting} & \textbf{Value} \\
\midrule
Task IDs & 106, 137, 145, 711, 1145 \\
Criteria evaluated & 76 total across 5 tasks \\
Resamples per config & 3 \\
Judge models & Claude Sonnet 4.5, Claude Sonnet 4.6, DeepSeek V3.2, Kimi K2.5 \\
Prompt versions & v1 (ref-guided classifier), v2 (ref-guided CoT), v3 (plain classifier), v4 (plain CoT) \\
Judge temperature & Model default (no explicit config) \\
Inference method & AWS Bedrock Converse API (live, not batch) \\
Total observations & 1,216 criterion-level verdicts \\
\bottomrule
\end{tabular}
\end{table}

The results of the stability experiment can be seen in Table~\ref{tab:stability}:

\begin{table}[H]
\centering
\caption{Stability by Model $\times$ Prompt (76 criteria, 3 resamples each).}
\label{tab:stability}
\small
\begin{tabular}{llcc}
\toprule
\textbf{Judge Model} & \textbf{Prompt} & \textbf{Stability \%} & \textbf{Flip Rate \%} \\
\midrule
Sonnet 4.6 & v1 (ref-guided) & 100.0\% & 0.0\% \\
Kimi K2.5 & v1 (ref-guided) & 98.7\% & 1.3\% \\
Sonnet 4.5 & v1 (ref-guided) & 97.4\% & 2.6\% \\
Sonnet 4.6 & v3 (plain) & 96.1\% & 3.9\% \\
DeepSeek V3.2 & v1 (ref-guided) & 94.7\% & 5.3\% \\
Sonnet 4.5 & v3 (plain) & 93.4\% & 6.6\% \\
Sonnet 4.6 & v2 (ref-guided CoT) & 92.1\% & 7.9\% \\
Sonnet 4.6 & v4 (plain CoT) & 90.8\% & 9.2\% \\
DeepSeek V3.2 & v3 (plain) & 89.5\% & 10.5\% \\
Sonnet 4.5 & v2 (ref-guided CoT) & 86.8\% & 13.2\% \\
Kimi K2.5 & v4 (plain CoT) & 85.5\% & 14.5\% \\
Sonnet 4.5 & v4 (plain CoT) & 84.2\% & 15.8\% \\
DeepSeek V3.2 & v2 (ref-guided CoT) & 80.3\% & 19.7\% \\
DeepSeek V3.2 & v4 (plain CoT) & 76.3\% & 23.7\% \\
Kimi K2.5 & v2 (CoT) & 72.4\% & 27.6\% \\
Kimi K2.5 & v3 (plain) & 72.4\% & 27.6\% \\
\bottomrule
\end{tabular}
\end{table}

Key findings:
\begin{itemize}
    \item \textbf{Most stable:} Sonnet 4.6 with v1 (reference-guided classifier) achieved 100\% stability---all 76 criterion verdicts identical across three independent runs.
    \item \textbf{CoT reduces stability:} Across all four models, CoT prompt versions averaged approximately 5 percentage points lower stability than their classifier counterparts (undesriable for a production scoring pipeline)
    \item \textbf{Model ranking for stability:} Sonnet 4.6 $>$ Sonnet 4.5 $>$ DeepSeek V3.2 $>$ Kimi K2.5.
\end{itemize}

Despite lower stability, the reasoning traces produced by v2 and v4 prompts provided qualitative insight into judge decision-making and failure modes. The 101 disagreement cases (criteria where the three runs did not unanimously agree) were qualitatively analysed and categorised into six failure modes shown in Table~\ref{tab:failure-modes}:

\begin{table}[H]
\centering
\caption{Failure mode categorization (101 flip cases).}
\label{tab:failure-modes}
\small
\begin{tabular}{clcc}
\toprule
\textbf{Category} & \textbf{Count} & \textbf{\% of flips} \\
\midrule
Threshold disagreement & 41 & 40.6\% \\
Criterion interpretation ambiguity & 21 & 20.8\% \\
Negation/polarity confusion & 11 & 10.9\% \\
Implicit vs.\ explicit requirement & 10 & 9.9\% \\
Scope of evaluation (meta-commentary) & 9 & 8.9\% \\
Audience/target confusion & 9 & 8.9\% \\
\bottomrule
\end{tabular}
\end{table}

Threshold disagreement dominated, accounting for 40.6\% of all flips. This category is inherent to any binary classification of continuous quality and is unlikely to be fully eliminated (similar borderline considerations are flagged by human raters in their optional free-text comments in the Practitioner Study Validation). Negation/polarity confusion (NPC) disproportionately affected DeepSeek V3.2 (5 of 11 NPC cases) and suggests a surface-form parsing limitation that could be addressed through prompt engineering.

\subsection{Judge Selection}
\label{app:judge-config}

Based on convergent evidence from Parts A, B, and the stability analysis, we selected \textbf{Claude Sonnet 4.6 with v1 (reference-guided classifier)} as the production judge. This configuration achieved the best balance across three dimensions:
\begin{enumerate}
    \item \textbf{Human alignment} (Part A): F1 = 0.936, $\kappa = 0.719$---second-best F1, only 0.6 points below DeepSeek V3.2.
    \item \textbf{Stability:} 100\% verdict consistency across re-runs---the only configuration with perfect stability.
    \item \textbf{Cost:} Approximately \$0.003 per task, making full-dataset scoring economically viable.
\end{enumerate}

\begin{table}[h]
\centering
\caption{Top 3 Judge Configurations Comparison.}
\label{tab:top3-judges}
\small
\begin{tabular}{lcccc}
\toprule
\textbf{Configuration} & \textbf{F1 (A)} & \textbf{Acc. (B)} & \textbf{Stability} & \textbf{Cost/Task} \\
\midrule
\textbf{Sonnet 4.6 v1} (selected) & 0.936 & 89.1\% & 100.0\% & $\sim$\$0.003 \\
DeepSeek V3.2 v1 & 0.942 & 86.2\% & 94.7\% & $\sim$\$0.002 \\
Sonnet 4.5 v1 & 0.925 & 87.4\% & 97.4\% & $\sim$\$0.004 \\
\bottomrule
\end{tabular}
\end{table}

\subsubsection{Cross-family judge agreement.}
\label{app:cross-family}
Because the production judge (Claude Sonnet 4.6) belongs to the same model family as the top-ranked benchmark entrant (Claude Opus 4.7), we quantified the extent to which the production verdicts could be reproduced by a judge from an unrelated model family. Using the shared 326-criterion subset scored by both the Claude Sonnet 4.6 (v1) and DeepSeek V3.2 (v1) judges on identical solver responses, the two judges agreed on the raw Pass/Fail verdict for 92.3\% of criteria, with Cohen's $\kappa = 0.764$ (substantial agreement). This is comparable to each judge's independent alignment with human raters (Table \ref{tab:judge-human-alignment}), indicating that the criterion-level verdicts are not an artefact of a single model family.

At the level of per-competency aggregate scores, Spearman's $\rho = 0.196$ ($p = 0.564$) and Kendall's $\tau = 0.147$ ($p = 0.532$) across the 11 competencies represented in the subset. The weak rank correlation is expected and does not indicate disagreement: aggregate competency pass-rates are tightly clustered near ceiling on this reference-guided subset, so small absolute differences produce unstable rank orderings. The high criterion-level agreement ($\kappa = 0.764$) is the more informative statistic for judge robustness.

Notwithstanding, to build on our small experiments here, a fuller multi-family, multi-prompt judge audit is left to future work (see Appendix~A).

\clearpage

\section{L2-Bench Results}
\label{app:results}

This appendix provides additional details on the L2-Bench scoring methodology and results.

\subsection{Scoring Formula}
\label{app:scoring-formula}

Task scores are computed from per-criterion binary Pass/Fail verdicts:

\begin{equation}
\text{task\_score} = \frac{\sum \text{passed\_weights}}{\sum \text{positive\_weights\_only}}
\end{equation}

\begin{itemize}
    \item \textbf{Positive criteria} (weight $> 0$): Passing adds the weight to the numerator.
    \item \textbf{Negative criteria} (weight $< 0$): If the undesirable behaviour is detected, the negative weight reduces the numerator. If absent, 0 is added---no penalty.
    \item \textbf{Denominator:} Only positive weights contribute (maximum achievable ignoring penalties).
    \item \textbf{Score range:} Scores can be negative when multiple penalties activate; clipped to 0 for reporting.
\end{itemize}

\subsection{Statistical Methods}
\label{app:ci-rank}

\paragraph{L2-Bench aggregate score standard error.} Following the recommendations of \citet{miller2024addingerrorbarsevals} for reliable benchmark evaluation, a model's overall L2-Bench score is treated as an estimate whose uncertainty is captured by its standard error. The 1{,}000 L2-Bench tasks are not exhaustive but are drawn from a hypothetical super-population of language-education tasks, and each task is scored over $k=3$ independent runs. The standard error therefore decomposes into two additive components---a between-task (super-population) variance and a within-task response/judge variance reduced by resampling:
\begin{equation}
\mathrm{SE}^2 = \frac{\operatorname{Var}(x) + \operatorname{E}\!\left[\sigma_i^2 / k\right]}{N}
\label{eq:score-se}
\end{equation}
where $\operatorname{Var}(x)$ is the variance of the task-level mean scores, $\sigma_i^2$ the within-task variance across the $k$ runs for task $i$, and $N = 1{,}000$. In practice $\operatorname{Var}(x)$ dominates (85--92\% of total variance), confirming that $k=3$ resamples adequately suppress within-task noise. We report a 95\% confidence interval using the $t$-distribution with $N-1$ degrees of freedom, $\bar{x} \pm t_{0.975,\,N-1}\,\mathrm{SE}$. These intervals capture sampling variability over tasks; they do not propagate judge- or criterion-level uncertainty, which we flag as future work (see Appendix~A).

\paragraph{Competency heterogeneity (ANOVA / Kruskal--Wallis).} To test whether a model's performance is uniform across the 12 competencies, we apply, per model, a one-way ANOVA and a non-parametric Kruskal--Wallis test over the competency-level scores, reporting $\eta^2$ as the effect size:
\begin{equation}
\eta^2 = \frac{SS_{\text{between}}}{SS_{\text{total}}}, \qquad
H = \frac{12}{n(n+1)}\sum_{g=1}^{G}\frac{R_g^2}{n_g} - 3(n+1)
\label{eq:anova-kw}
\end{equation}
where $SS_{\text{between}}$ and $SS_{\text{total}}$ are the between-competency and total sums of squares, and $H$ is the Kruskal--Wallis statistic over $G$ competency groups with rank sums $R_g$ and sizes $n_g$. All nine models reject the null of uniform competency performance ($p < 0.001$ on both tests; Table~\ref{tab:anova_kw_models}), confirming that competency profile is a genuine and model-specific source of variation rather than noise.

\paragraph{Rank stability across variants.} We assess how sensitive the leaderboard ordering is to the choice of scoring variant using Kendall's $\tau$ between the plain ranking and each alternative variant:
\begin{equation}
\tau = \frac{n_c - n_d}{\tfrac{1}{2}\,m(m-1)}
\label{eq:kendall-tau}
\end{equation}
where $n_c$ and $n_d$ are the numbers of concordant and discordant model pairs and $m=9$ models. Rankings are highly stable: $\tau = 1.00$ for both the hard-items and validated-subset variants and $\tau = 0.89$ for the length-adjusted (verbosity) variant (Spearman's $\rho = 1.00$, $1.00$, and $0.95$ respectively). All $\tau$ values exceed the $0.7$ threshold cited in the main text, and the top tier is preserved under every variant, so the headline ordering is not an artefact of the specific aggregation rule. The one exception---GPT~5.4 dropping under length adjustment---is discussed in Appendix~I.4 below.

\subsection{Model Selection}
\label{app:model-selection}

For initial L2-Bench results, we selected nine frontier, mid-sized, and small models for general-purpose reasoning rather than specialised capabilities such as coding. To assess out-of-the-box behaviour that reflects the baseline that standard enterprise users would encounter, all models were queried through official cloud APIs (Microsoft Azure, Google Vertex AI, and AWS Bedrock) using each provider's default inference configuration. Where documented, the known default parameters were: Temperature~$=1.0$, Top-$p=0.95$, and Top-$k=64$ for Vertex's Gemini models; Temperature~$=1.0$ for the Azure OpenAI and DeepSeek APIs. All 27{,}000 scored task-response runs (9 models $\times$ 1{,}000 tasks $\times$ 3 runs) will be open-sourced alongside the benchmark (see Appendix~B). Model coverage is a limitation of this release and will expand in future rounds (see Appendix~A).

\subsection{Score Variants}
\label{app:score-variants}

Beyond the plain leaderboard, we report four score variants to probe robustness in Table~\ref{tab:leaderboard-variants}.

\paragraph{Length-adjusted (verbosity) scores.} The verbosity variant penalises response length so that models are rewarded for pedagogical quality \emph{per token}. Under this adjustment GPT~5.4 drops from 2nd to 4th (below Gemini 3.1 Pro and Gemini 3 Flash), indicating that its plain-score standing is partly supported by longer responses that may be more burdensome for adopters to consume.

\paragraph{L2-Bench Hard.} The hard variant restricts scoring to the 267 tasks on which the top-three models all scored below 80\%, isolating the most demanding items. Absolute scores fall by roughly 12--15 percentage points across the board, but the ordering is unchanged ($\tau = 1.00$ vs.\ plain), showing that the leaderboard's separation of tiers is driven by genuine difficulty rather than easy items alone.

\paragraph{Validated scores.} The validated variant restricts scoring to the 504 tasks whose rubrics were reviewed and confirmed by expert practitioners in the validation study (Appendix~G), so that the leaderboard reflects only items with externally verified construct validity. The ordering is again unchanged ($\tau = 1.00$ vs.\ plain), indicating that model rankings do not depend on the unvalidated remainder of the dataset.

\begin{table}[h]
\centering
\caption{Leaderboard score variants (\% overall). Plain: full 1{,}000-task leaderboard; Hard: 267 hardest tasks; Verbosity: length-adjusted; Validated: 504 practitioner-validated tasks. Ranks in parentheses.}
\label{tab:leaderboard-variants}
\small
\begin{tabular}{lcccc}
\toprule
\textbf{Model} & \textbf{Plain} & \textbf{Hard} & \textbf{Verbosity} & \textbf{Validated} \\
\midrule
Opus 4.7        & 85.5 (1) & 73.4 (1) & 84.0 (1) & 87.3 (1) \\
GPT 5.4         & 84.1 (2) & 71.6 (2) & 80.9 (4) & 85.4 (2) \\
Gemini 3.1 Pro      & 83.4 (3) & 69.9 (3) & 83.7 (2) & 85.2 (3) \\
Gemini 3 Flash    & 80.7 (4) & 68.5 (4) & 81.7 (3) & 82.1 (4) \\
DeepSeek V3.2   & 80.2 (5) & 67.1 (5) & 80.6 (5) & 81.5 (5) \\
Kimi K2.5       & 79.1 (6) & 66.6 (6) & 80.1 (6) & 80.7 (6) \\
Haiku 4.5       & 78.8 (7) & 65.4 (7) & 77.1 (7) & 80.5 (7) \\
Qwen3 32B       & 65.8 (8) & 53.3 (8) & 66.3 (8) & 66.0 (8) \\
Magistral Small & 50.7 (9) & 40.1 (9) & 53.3 (9) & 49.5 (9) \\
\bottomrule
\end{tabular}
\end{table}

\subsection{Performance Across Task Contexts}
\label{app:task-contexts}

This appendix subsection details L2-Bench performance across various task contexts (for task distribution volumes across these contexts, expressed as shares of the full dataset, see Appendix~F.4).

Table~\ref{tab:role_performance} shows model performance across role-based evaluation dimensions. To avoid over-interpreting low-sample cells, we omit role combinations (\textbf{Assessment \& Teacher}, \textbf{Teacher \& Learner}, and \textbf{Curriculum \& Teacher}).

\begin{table}[h]
\centering
\small
\caption{Model performance by stakeholder persona / role of asker.}
\label{tab:role_performance}
\begin{tabular}{lccccc}
\toprule
\textbf{Model} & \textbf{Assess. Dev.} & \textbf{Curric. Design} & \textbf{Guide/Trainer} & \textbf{Learner} & \textbf{Teacher} \\
\midrule
Opus 4.7        & 81.3 & 87.4 & 91.7 & 81.5 & 85.7 \\
GPT 5.4         & 79.5 & 87.0 & 90.7 & 78.2 & 84.4 \\
Gemini 3.1 Pro      & 76.6 & 81.6 & 89.0 & 82.9 & 83.8 \\
Gemini 3 Flash    & 78.5 & 76.2 & 86.1 & 80.9 & 81.0 \\
DeepSeek V3.2   & 69.8 & 79.4 & 87.0 & 76.8 & 80.9 \\
Kimi K2.5       & 71.4 & 80.8 & 86.3 & 77.6 & 79.1 \\
Haiku 4.5       & 65.9 & 79.9 & 88.0 & 76.2 & 79.0 \\
Qwen3 32B       & 37.9 & 55.3 & 72.8 & 72.0 & 67.1 \\
Magistral Small & 31.4 & 39.6 & 57.3 & 61.2 & 51.1 \\
\bottomrule
\end{tabular}
\end{table}









\clearpage

\end{document}